\pdfoutput=1

\documentclass[11pt]{article}

\usepackage[final]{acl}

\usepackage{times}
\usepackage{latexsym}

\usepackage[T1]{fontenc}

\usepackage[utf8]{inputenc}

\usepackage{microtype}

\usepackage{inconsolata}

\usepackage{graphicx}

\usepackage{color}
\usepackage{soul}
\usepackage{adjustbox}
\usepackage{array} 
\usepackage{tabularx}
\usepackage{booktabs}
\usepackage{hyperref}
\usepackage{makecell}
\usepackage{multirow}
\usepackage{subcaption}
\usepackage[ruled,vlined,linesnumbered]{algorithm2e}
\usepackage{float}
\usepackage{amsmath}
\usepackage[dvipsnames]{xcolor}

\usepackage{amssymb}
\usepackage{pifont}
\newcommand{\cmark}{\ding{51}}%
\newcommand{\xmark}{\ding{55}}%

\title{BRANCH: Bypassing Multi-Scanner AI Guardrails}

\author{
 \textbf{William Hackett\textsuperscript{1,2}}
 \textbf{Peter Garraghan\textsuperscript{1,2}}
\\
\\
 \textsuperscript{1}Mindgard,
 \textsuperscript{2}Lancaster University
\\
 \small{ 
   {\{william.hackett, peter\}@mindgard.ai}
 }
 \\
}

\DeclareMathOperator*{\argmax}{arg\,max}

\begin{document}
\maketitle

\begin{abstract}
AI systems increasingly rely on Large Language Models (LLMs) as core reasoning engines, making them targets for prompt injection and jailbreaks. Guardrails monitor and validate model inputs and outputs, yet their isolated, task-focused detection leaves gaps in their classification making them susceptible to bypasses. In response, \textit{guardrail systems} formed by multiple \textit{scanners} have emerged that collaboratively detect different types of malicious instructions, whereby shared latent representations across classification boundaries render established bypassing techniques ineffective. We propose \textit{BRANCH}, a bypassing methodology designed for multi-scanner guardrail systems. Our method leverages a branching tree search approach that dynamically applies adversarial perturbation against individual scanners, with subsequent perturbation optimization and technique selection based on overall improvement across all guardrail system scanners, effectively decoupling bypass evaluation from attack signal optimization. Our findings demonstrate that BRANCH achieves 100\% attack success rate across 6 guardrail systems in 120 scenarios with 72\% fewer queries and 4.5x reduced wall-clock time compared to established techniques, while preserving semantic meaning within the bypass. We also show how bypasses generated by BRANCH transfer to 29 unseen guardrails, including 8 commercial black-box guardrails, improving attack success in some cases up to 100\% with no additional optimization.


\end{abstract}

\section{Introduction}
\label{introduction} 

The adoption of \textit{Generative Artificial Intelligence (GenAI)} has led to the rapid proliferation of \textit{AI systems} that use these models to perform complex tasks \cite{electronics14132717, shethiya2023llm}. Subsequently, AI systems have become targets for adversarial prompting techniques such as prompt injection and jailbreaking that attackers use to exfiltrate sensitive data, manipulate model behavior, or escape trained safety alignment \cite{wu2024new}.

To mitigate these attacks, \textit{guardrails} were created to monitor and validate inputs and outputs to GenAI models within an AI system. Guardrails typically use a range of detection techniques such as text-classification models, regular expression, and pattern analysis designed to detect specific conent categories including toxicity, prompt injection, and jailbreaks \cite{ayyamperumal2024current, dong2024safeguardinglargelanguagemodels}. An individual single-purpose guardrail, which we refer to as a \textit{scanner} provides narrow, single-category classification that is susceptible to the evasion techniques established against text classifiers such as character injection and word substitution \cite{bassani2025guardrail, hackettbypassing}. In response, \textit{guardrail systems} have emerged as the more sophisticated solution, whereby an ensemble of scanners are combined to detect malicious content across categories, covering blindspots in detection capabilities. This forces the attacker to bypass every scanner within the system simultaneously in order to succeed \cite{llmguard2025,vijildome2024,llamafirewall}.

Although guardrail systems have become commonplace within production AI systems, no research has evaluated the effectiveness of current attacks against multiple scanners simultaneously. We hypothesize that the presence of multiple scanners creates a multi-objective optimization problem for attackers, whereby the \textit{shared latent representations} across scanners constrain established bypassing techniques in overlapping decision boundaries \cite{miettinen1999nonlinear, demontis2019adversarialattackstransferexplaining}. As a result, current techniques either fail to generate a bypass that evades all guardrails at once, or succeed only at substantial computational cost.

In this paper, we propose \textit{BRANCH}, a bypassing methodology designed for bypassing guardrail systems. BRANCH leverages a branching tree search that chains multiple perturbation techniques together, isolating bypassing to individual scanners, while selecting branches which gradually bypass all scanners within a guardrail system simultaneously. Our approach is applicable to any malicious instruction while being perturbation agnostic, enabling the generation of bypasses against guardrail systems comprising of multiple scanners. Furthermore, we demonstrate that bypasses generated by BRANCH also transfer to a large set of guardrails unseen during generation. To our knowledge, our work is the first approach to reliably bypass more sophisticated guardrail systems as a whole, instead of individual guardrail scanners.

We make the following contributions:
\begin{enumerate}
    \item We investigate the effectiveness of current single-objective evasion techniques against multi-scanner guardrail systems, demonstrating their challenges and limitations when operating within multi-objective bypass scenarios, achieving lower success at substantive computational cost (Section~\ref{multi_objective_optimisation_problem}).
    \item We propose a novel multi-objective guardrail bypassing methodology called \textit{BRANCH} to address the challenges of established single-objective evasion techniques, and demonstrate its ability to achieve a 100\% bypass success rate across 6 target guardrail systems using 72\% fewer queries and 4.5x less wall-clock time compared to the State-of-the-Art, while preserving semantic content comparable to established techniques (Section~\ref{result:bypass_effectiveness}, Section~\ref{result:generated_bypass_similarity}).
    \item We demonstrate that bypasses generated by BRANCH successfully transfer and increased Attack Success Rate by an average of 32.9 percentage points across 29 unseen production guardrails without any adversarial perturbation, including 8 commercially offered black-box guardrail systems. (Section~\ref{result:transferability}).
\end{enumerate}

\section{Background}
\label{background}

\subsection{Guardrails}
\label{guardrails}

Guardrails are predictive systems embedded within AI systems which monitor the inputs and outputs of GenAI models, enabling an AI system to react if malicious content is detected \cite{ayyamperumal2024current}. We categorize guardrails into two types: \textit{scanners}, which are individual, single-purpose guardrails designed to detect a single category of content, and \textit{guardrail systems}, which are ensembles of scanners that combine their outputs to make a single classification decision.

\textbf{Guardrail Scanners.} Designed to detect a single category of content such as toxicity, prompt injections, or jailbreaks. A scanner may be implemented through a range of detection techniques. The most common are AI-based, using text-classification models pre-trained to detect a specific category of content, while others leverge regular expression and pattern matching, or LLM-as-a-judge judge \cite{nemoguardrails, llmjudges}. Text-classification has been widely adopted because it can be pre-trained on datasets containing examples of the content it is designed to detect, using architectures such as transformer-based models (e.g. BERT) that are fine-tuned to classify inputs as malicious or non-malicious based on the linguistic features and patterns \cite{PIGuard,dubey2024llama3herdmodels,microsoft2024contentfilter}. This allows a scanner to generalize beyond its training examples to detect a wide range of unseen malicious inputs. However, because each scanner is trained to detect a single category of content, its classification is isolated and task-focused, providing no coverage of the categories it was not trained on \cite{hackettbypassing}.

\textbf{Guardrail Systems.} These systems address scanner limitations by leveraging multiple scanners designed for different categories in tandem to create a single classification decision to provide more comprehensive coverage across threat categories \cite{llmguard2025,vijildome2024,llamafirewall}. This approach prevents attackers from exploiting weaknesses within the classification capabilities of a single scanner as the decision boundaries in a guardrail system are more complex. Production guardrail systems offered by major vendors such as OpenAI, Microsoft, and Google combine multiple scanners to detect malicious content across categories \cite{microsoft2024contentfilter, zeng2024shieldgemmagenerativeaicontent, markov2023holisticapproachundesiredcontent}.

\subsection{Bypassing Techniques}
\label{bypassing_guardrails}

\begin{figure*}[tbp]
    \centering
    \begin{subfigure}[b]{0.49\linewidth}
        \centering
        \includegraphics[width=\linewidth]{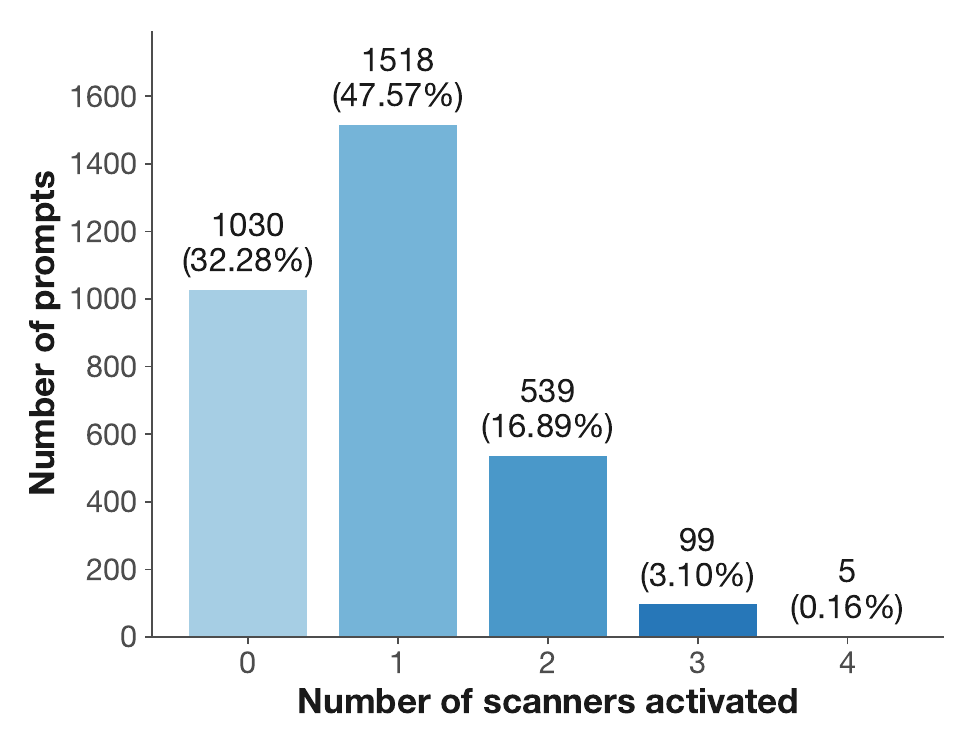}
        \caption{Total Scanner Activations.}
        \label{fig:scanner_activation_histogram}
    \end{subfigure}
    \hfill
    \begin{subfigure}[b]{0.49\linewidth}
        \centering
        \includegraphics[width=\linewidth]{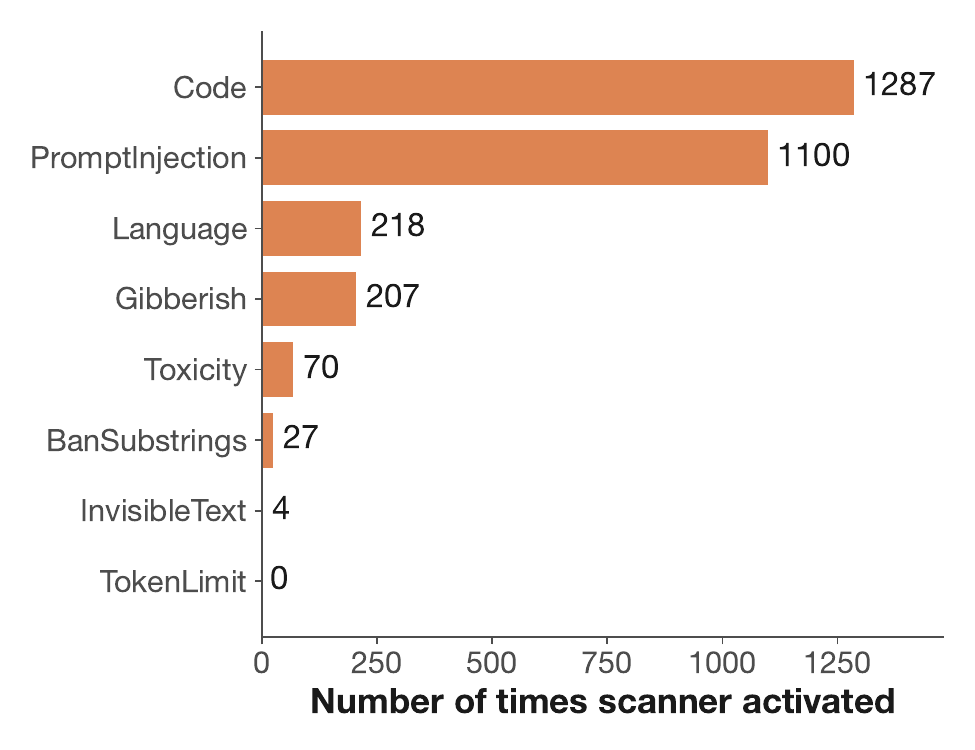}
        \caption{Activations Per Scanner.}
        \label{fig:scanner_activation_counts}
    \end{subfigure}
    \caption{\textbf{Static Scanner Activations.} The number of scanners activating, and their distribution, across 8 scanners within LLM Guard during inference of 3,191 prompt injection and jailbreak samples from PIGuard~\cite{PIGuard}.}
    \label{fig:overlapping_decision_boundaries}
\end{figure*}

The goal of an attacker targeting a guardrail system is to generate a \textit{bypass}, an input that evades the target AI system so that it is classified as non-malicious \cite{hackettbypassing}. Bypasses originate from \textit{Adversarial Machine Learning (AML) evasion}, that generates adversarial examples against image and text classifiers.

\textbf{Adversarial ML Evasion.} These techniques were adapted to guardrails due to their reliance upon ML models (such as text classifiers) to detect malicious content. Bypassing techniques differ in how they generate perturbations. For example, genetic algorithms can be used to evolve the input \cite{alzantot}, BERT-based masking to identify and replace most important tokens  \cite{bae,bertattack}, and substituting characters with homoglyphs or intentional typos \cite{textbugger, textfooler}. Which tokens that the technique modifies is determined by \textit{word importance ranking}, which scores each word by the change in the guardrail's confidence when that word is removed and prioritizes higher-impact words for modification \cite{textfooler,pwws}. This enables an attacker to target features within the input which the target directly relies upon for classification, allowing these areas to be modified more effectively \cite{evasion_attacks}.

\textbf{Perturbation Methods.} A \textit{perturbation} is a modification applied to the input to generate a bypass \cite{morris2020textattack}. Perturbation methods differ in the granularity at which they are applied, which we separate into two categories \cite{bae,bertattack,textbugger,deepwordbug}. \textit{Character injection} inserts characters into the input to evade detection, either invisible characters such as zero-width spaces, or characters that are visually similar to the originals \cite{deepwordbug}. Although these perturbations are perceptible to humans, GenAI models remain capable of interpreting the underlying obfuscated or injected input. \textit{Word substitution}, by contrast, replaces words in the input with synonyms or similar alternatives, generating substitutes through methods such as BERT-based masking \cite{bae,bertattack}, or word search algorithms \cite{bird-loper-2004-nltk}.

\section{The Challenges of Established Bypassing Techniques Against Guardrail Systems}
\label{multi_objective_optimisation_problem}

As guardrail systems have more complex decision boundaries compared to single scanners, single-objective evasion techniques struggle when attempting to successfully bypass multiple scanners simultaneously. We attribute the challenges of these techniques to two contributing factors: \textit{overlapping decision boundaries} between guardrails that share latent representations, and \textit{scanner interference}, where a single-objective technique must collapse the per-scanner confidence scores into one scalar, resulting in treating all scanners as a single target, and causing guardrails which were not actively blocking to block when perturbations are applied to evade another.

\subsection{Overlapping Decision Boundaries}
\label{overlapping_decision_boundaries}

\begin{figure*}[h]
    \centering
    \includegraphics[width=\linewidth]{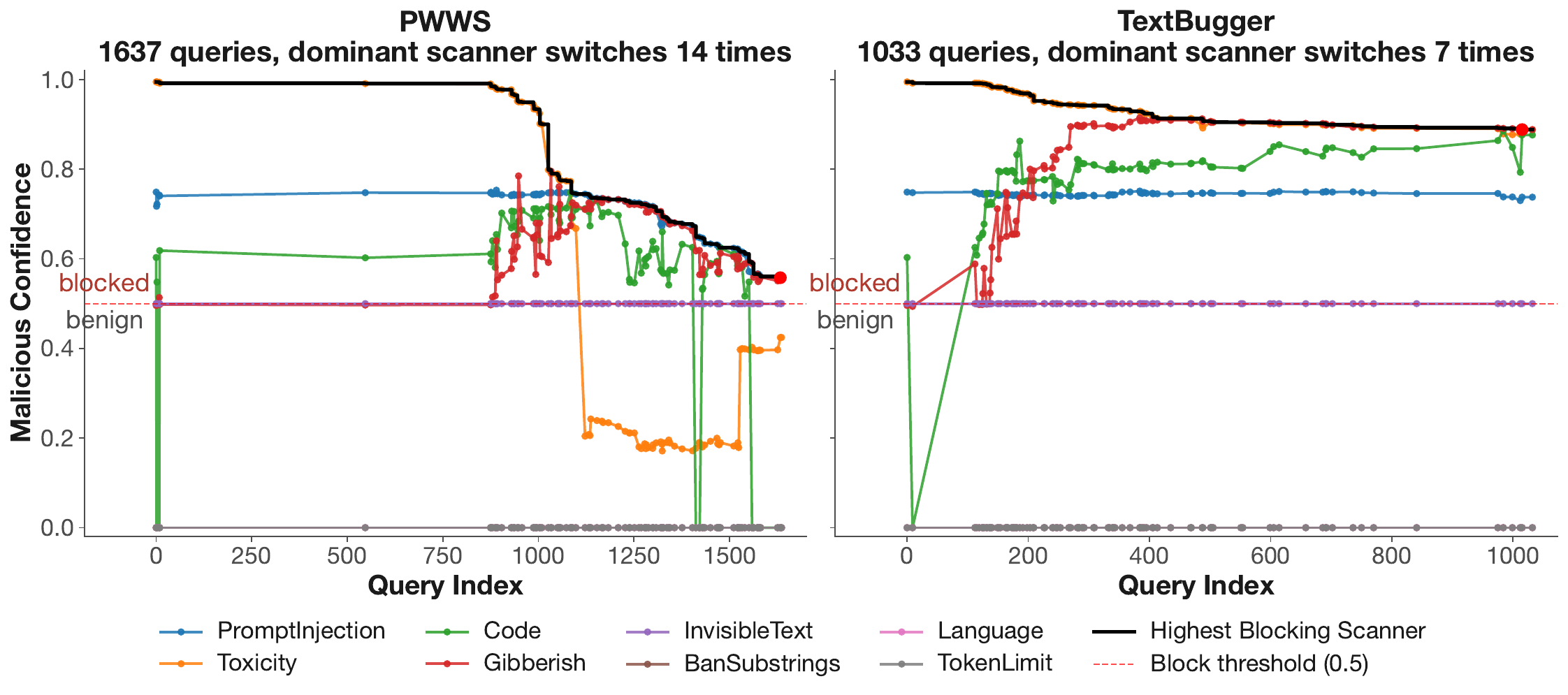}
    \caption{\textbf{Scanner Displacement.} Scanner confidences tracked across two single-objective techniques, \texttt{PWWS} and \texttt{TextBugger} showing the currently highest blocking scanner and variance of other scanners during perturbation.}
    \label{fig:scanner_displacement}
\end{figure*}

The shared latent representations of scanners within a guardrail system can lead to overlapping decision boundaries, where the same input may be classified as malicious by multiple scanners due to similar features or patterns associated with the content each scanner is designed to detect. As scanners share these latent features, perturbations that reduce the confidence of one scanner by moving the input across a decision boundary can simultaneously increase another scanner's confidence score, since their correlated feature spaces can act in opposition \cite{demontis2019adversarialattackstransferexplaining}.

\textbf{Shared Latent Representations.} If we consider a guardrail system comprising two scanners -- prompt injection and jailbreak detection -- both scanners may contain overlapping decision boundaries due to similar features typically associated with both prompt injections and jailbreaks (i.e. shared phrase and sentence structures that attempt to manipulate the model, or specific malicious payloads) \cite{rossi2024earlycategorizationpromptinjection, GENG2026}. An example of this overlap exists within AML literature, where prompt injection and jailbreaks are often conflated or misinterpreted \cite{rossi2024earlycategorizationpromptinjection}. Furthermore, prompt injection and jailbreaks are often embedded with other malicious content, such as toxicity or code, which can increase such overlap further \cite{PIGuard}.

\textbf{Static Scanner Activation.} To quantify this overlap, we construct a dataset of 3,191 prompt injection and jailbreak samples sourced from PIGuard \cite{PIGuard}, and record the classification score of every scanner contained within the guardrail system LLM Guard for each sample. Figure~\ref{fig:scanner_activation_histogram} shows the resulting activation distribution, where 32.28\% of samples activate no scanner, 47.57\% activate exactly one, and the remaining 20.15\% activate two or more. Figure~\ref{fig:scanner_activation_counts} demonstrates that 7 of the 8 scanners activate at least once across the dataset, with Code accounting for the highest number activations at 1,287, followed by PromptInjection at 1,100. This spread indicates that despite every sample being labelled as a prompt injection or jailbreak, the samples carry features that scanners trained for other categories detect. While only 20.15\% of samples activate two or more scanners simultaneously, it is worth noting that these are static prompts measured before any perturbation is applied. Therefore, when an attacker applies perturbation to an input, it is likely that another scanner's confidence score increases above the designated threshold and activates.

\subsection{Scanner Interference}
\label{scanner_fighting}

When a guardrail system contains multiple scanners, each query produces a separate confidence score per scanner rather than a single value. Current evasion techniques, such as \texttt{PWWS} and \texttt{TextBugger}, are designed around a single-objective, meaning they require a single confidence signal to guide the bypass search \cite{morris2020textattack}. Applying these techniques to a guardrail system therefore requires an aggregation function that collapses the per-scanner confidences into a singular scalar, all of which introduce their own trade-offs \cite{miettinen1999nonlinear}. However, collapsing multiple scanner confidence scores into a single scalar results in \textit{scanner interference}, which from the perspective of the adversarial technique, this results in behavior of a scanner is indistinguishable from another, limiting the ability for a bypassing technique to optimize adversarial samples based upon feedback of all scanners.

\textbf{Scanner Displacement.} We demonstrate the problem of scanner interference by tracing every classification query made by two bypassing techniques, PWWS and TextBugger, when applied to one seed prompt that performs adversarial perturbation in response to classification scoring against LLM Guard. Figure~\ref{fig:scanner_displacement} shows that both techniques repeatedly suppress the currently dominating scanner only for a different scanner, with the dominant scanner switching 14 times over PWWS's 1,637 queries and 7 times over TextBugger's 1,033 queries leading the techniques to optimize against only the dominant scanner in those scenarios. This requires the techniques to navigate a more sophisticated decision boundary, learning to generate perturbations which appear to be effective against the target until another scanner becomes the dominant blocking scanner. Furthermore, these techniques typically perform word importance ranking before entering the perturbation stage, meaning the computed importance ranking is bespoke to the initial dominant scanner impacting effectiveness of the perturbations, and not representative of confidence across the entire guardrail system \cite{pwws,textbugger}. This interference therefore influences the effectiveness of these techniques, limiting their ability to generate bypasses when facing a multi-objective bypass scenario. 

\section{BRANCH}
\label{branch}

Addressing the challenges of single-objective bypassing techniques discussed in Section~\ref{multi_objective_optimisation_problem}, we propose \textit{BRANCH}, a multi-objective bypassing methodology that uses a branching tree search to bypass multiple scanners within a guardrail system. Each node in the search tree isolates a single \textit{target scanner} that represents the blocking scanner with the highest confidence, and focuses perturbation techniques to target the scanner irrespective of changes to other scanner confidence scores, giving every perturbation a well-defined single objective. Once a perturbation technique completes, the resulting adversarial sample is considered a \textit{candidate}, which are scored on the \textit{total blocked confidence} summed across every scanner. This enables BRANCH to address scanner interference and overlapping decision boundaries by reducing the multi-objective bypass scenario into a single-objective, targeting only one individual scanner at a time, while rewarding perturbations that gradually bypass all scanners within the target. Therefore, by expanding the node with the lowest total blocked confidence and seeding each technique with the output of the last, BRANCH accumulates perturbations that lower confidence across the guardrail system as a whole rather than attempting to suppress all scanners within one perturbation technique. 

\subsection{Threat Model}

We consider an adversary whose goal is to generate a \textit{bypass} -- a prompt or instruction that successfully evades all scanners contained within a target guardrail system simultaneously, while preserving the malicious intent of the original seed prompt.

\textbf{Access.} We assume a \textit{black-box} access model whereby the adversary can submit arbitrary prompts to the target guardrail system and observe the per-scanner blocking decision and associated confidence score for each query, yet has no access to model weights, gradients, internal architectures, or training data. This access model is representative of established bypassing literature where an attacker can access confidence scores by locally running the guardrail system, or through guardrail vendor APIs \cite{hackettbypassing, textbugger, textfooler, zhou2026promptoverflowguardrailinspects}.

\textbf{Knowledge.} The adversary requires knowledge of the blocking decision and confidence scores returned by a specific scanner and overall guardrail system. Knowledge obtained via local access to the guardrail system, or some cases moderation APIs \cite{hackettbypassing, markov2023holistic}. The adversary does not need any knowledge about the AI model, the training data, or AI system the guardrail system is operating within. 

\textbf{Capability.} We assume the adversary is not constrained by a budget or rate limited. Our approach can be leveraged against open-source or self-hosted guardrails without vendor visibility, with successful bypasses generated subsequently applied to production guardrail systems (including unseen types). We report two metrics, \textit{query} and \textit{wall-clock} cost as measures of attack efficiency.

\subsection{Preliminaries}

\textbf{Notation.} Let $G$ be our target guardrail system comprising $k$ scanners, $G = \{s_1, \ldots, s_k\}$, and $P$ the original seed prompt considered for perturbation. Each scanner $s \in G$ produces a binary blocking decision $b$, along with a confidence score $c$, for an input prompt $p$.
\begin{equation}
b_s(p) \in \{0, 1\}, \quad c_s(p) \in [0, 1]
\end{equation}
where $b_s(p) = 1$ indicates that scanner $s$ blocks prompt $p$, and $c_s(p)$ is the associated with classification confidence. A prompt $p$ is \textit{blocked} by the guardrail system if one or more scanner block due to exceeding the configured scanner confidence score threshold, and \textit{evaded} if no scanner blocks it.
\begin{equation}
\begin{aligned}
\text{blocked}(p) &= \left[\forall s \in G : b_s(p) = 1\right] \\
\text{evaded}(p) &= \left[\forall s \in G : b_s(p) = 0\right]
\end{aligned}
\end{equation}

\textbf{Total Blocked Confidence.} BRANCH tracks evasion progress against a guardrail system by measuring the total summed confidence scores across blocking scanners.
\begin{equation}
C(p) = \sum_{s \in G} b_s(p) \cdot c_s(p)
\end{equation}
where $C(p)$ is the total summed confidence of currently blocking scanners in the guardrail system.

\textbf{Target Scanner.} The target scanner for $p$ is the blocking scanner returning the highest confidence.
\begin{equation}
\sigma^*(p) = \argmax_{s \in G\,:\,b_s(p) = 1} c_s(p)
\end{equation}
where $\sigma^*(p)$ represents the current target scanner with the highest confidence across all blocking scanners within the guardrail system.

\textbf{Tree Structure.} BRANCH maintains a search tree $T = (V, E)$ beginning with $P$ where each node $v \in V$ holds a prompt $p_v$, and each edge in $E$ records the node a candidate originated from. Each node contains its own target scanner, $\sigma^*(p_v)$, total blocked confidence $C(p_v)$, and depth $d_v$ with two optional constraints: $D_{\max}$ for maximum depth, and $N_{\max}$ for maximum nodes expanded.

\textbf{Bypass Candidate Set.} BRANCH runs an ensemble of evasion techniques to generate potential bypasses which are leveraged as a set of candidates $\mathcal{C}_v = \{q_1, \ldots, q_m\}$, where each $q_i$ is a modified variant of $p_v$ generated against the target scanner.

\begin{algorithm}
\footnotesize
\DontPrintSemicolon
\LinesNumbered
\caption{BRANCH algorithm}
\label{alg:branch}
\KwIn{initial seed prompt $P$; guardrail system $G$; perturbation techniques $\mathcal{T}$; node budget $N_{\max}$; max depth $D_{\max}$}
\KwOut{set of evaded prompts $P^{\ast}$}
$P^{\ast} \gets \emptyset$\;
evaluate $P$ against $G$\tcp{gather baseline}
$V_{\text{pending}} \gets \{\text{make\_root}(P) \mid P\ \text{blocked}\}$\;
$n \gets 0$\;
\BlankLine
\While{$V_{\text{pending}} \neq \emptyset$ \textbf{and} $n < N_{\max}$}{
    $v^{\ast} \gets \arg\min_{v \in V_{\text{pending}}} C(p_v)$ \tcp{select node based on $C(p_v)$}
    $V_{\text{pending}} \gets V_{\text{pending}} \setminus \{v^{\ast}\}$\;
    $n \gets n + 1$\;
    \BlankLine
    \tcc{Phase 1: Bypass Target Scanner}
    $\mathcal{C}_{v^{\ast}} \gets \textsc{Bypass}(p_{v^{\ast}}, \sigma^*(p_{v^{\ast}}), \mathcal{T})$\ \tcp{bypass job}
    \BlankLine
    \tcc{Phase 2: Validate}
    \ForAll{$q_i \in \mathcal{C}_{v^{\ast}}$}{
        evaluate $q_i$ against guardrail system $G$\;
        \If{$\text{evaded}(q_i)$}{
            $P^{\ast} \gets P^{\ast} \cup \{q_i\}$\;
        }
    }
    \If{$P^{\ast} \neq \emptyset$}{
        \Return{$P^{\ast}$} \tcp{early exit on bypass}
    }
    \BlankLine
    \tcc{Phase 3: Grow}
    \If{$d_{v^{\ast}} < D_{\max}$}{
        \ForAll{blocked $q_i \in \mathcal{C}_{v^{\ast}}$ \textbf{and} not in tree}{
            $v' \gets (q_i,\, d_{v^{\ast}} + 1)$\;
            $V_{\text{pending}} \gets V_{\text{pending}} \cup \{v'\}$;\quad $E \gets E \cup \{(v^{\ast}, v')\}$\;
        }
    }
}
\BlankLine
\Return{$P^{\ast}$}\;
\end{algorithm}

\subsection{Methodology}

\begin{figure*}
    \centering
    \includegraphics[width=\linewidth]{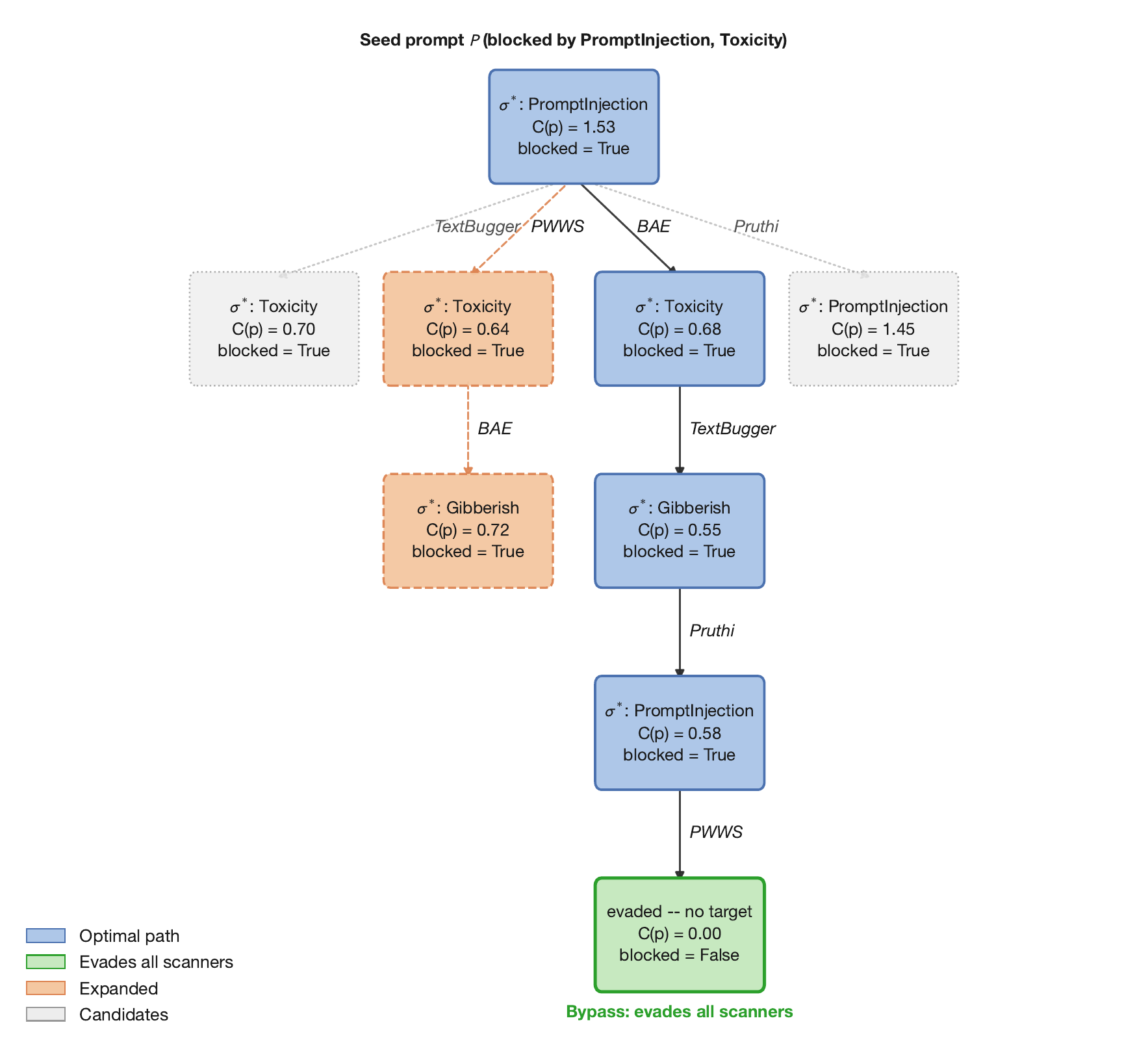}
    \caption{\textbf{BRANCH Methodology.} Example bypass against a guardrail system with three scanners: Prompt Injection, Toxicity, and Gibberish, using four perturbation techniques BAE, TextBugger, Pruthi, and PWWS. BRANCH selects the node with the lowest total blocked confidence ($(C(p)$), perturbs it with one or more techniques, and then validates candidates against the full guardrail system. The process is repeated until a bypass is found.}
    \label{fig:branch_tree_example}
\end{figure*}

BRANCH shown in Algorithm~\ref{alg:branch}, addresses the challenges that established techniques have against guardrail systems by focusing each node's search to a single scanner, rather than optimizing all scanners simultaneously. At each step, BRANCH selects the pending node with the lowest total blocked confidence across all scanners, prioritizing the prompt closest to evasion the the scanner returning the lowest confidence score, and then processes it in three phases. We refer to a branch which leads to a successful bypass as an \textit{optimal path}. We outline an example of this process in Figure~\ref{fig:branch_tree_example}.

\textbf{Phase 1: Bypass.} Generates prompt candidates modified through adversarial perturbation to degrade the performance of, or fully bypass, a specific target scanner. Perturbations are handled by an ensemble of techniques, which we term a \textit{bypass job}. As BRANCH is agnostic to the evasion technique, any collection of perturbation methods can be used. Furthermore, isolating to a single scanner enables single-objective perturbation techniques to operate within their limitations.

\textbf{Phase 2: Validate.} Focuses on validating generated candidates against the guardrail system. This phase ensures BRANCH considers the multi-objective nature of bypassing every scanner in a guardrail system. A candidate which triggers the guardrail system is recorded as blocked, whereas any candidates that are not blocked are reported as bypasses that will result in the search subsequently terminating.

\textbf{Phase 3: Grow.} Responsible for expanding the search tree, every candidate prompt produced which does not already exist within the tree, and is blocked by the guardrail system becomes a child node at that specific depth. Each node contains the candidate prompt it should perturbed, and the target scanner which currently has the highest confidence with that candidate. Candidates are then utilized as the input to the next node within the tree, enabling techniques to be chained together, and therefore exploring a more sophisticated search space.

\section{Experimental Setup}
\label{branch:experiments}

\textbf{Evaluation approaches.} We compare BRANCH against 11 prominent guardrail bypassing techniques (see Appendix Table \ref{tab:bypass_techniques}), selected based upon their diversity in perturbation methodologies \cite{morris2020textattack}. We evaluate three approaches. \textit{Individual Techniques}: Each individual technique applies adversarial perturbation to a seed prompt to measure the effectiveness of a single isolated bypass. \textit{Exhaustive}: All 11 techniques are ran, whereby a successful bypass is determined in the event of \emph{any} technique succeeding. \textit{BRANCH}: Our approach, which chains techniques across the search tree so that the output of one seeds the next. The query budgets differ across approaches, where individual techniques and Exhaustive utilize an unbounded query budget, terminating only when the technique exhausts its search, while BRANCH uses a maximum of 5,000 queries per node, a node budget $N_{\max} = 120$, and maximum depth $D_{\max} = 120$, selected to evaluate BRANCH's effectiveness and efficiency across both depth and breadth of the search tree. 

\begin{table}
\centering
\small
\setlength{\tabcolsep}{6pt}
\renewcommand{\arraystretch}{0.8}
\caption{\textbf{Target Guardrail Systems}. Guardrails targeted in this work, including total number of scanners.}
\label{tab:targets}
\begin{tabular}{@{}lc@{}}
\toprule
\textbf{Guardrail System} & \makecell{\textbf{Number of}\\\textbf{Scanners}} \\
\midrule
LLM Guard \cite{llmguard2025} & 8 \\
Vijil Dome \cite{vijildome2024} & 3 \\
Guardrails AI \cite{guardrailsai} & 3 \\
Samsung Guard \cite{SGuard-v1} & 2 \\
LlamaFirewall \cite{llamafirewall} & 2 \\
Ensemble Guardrail (Custom Built) & 3 \\
\bottomrule
\end{tabular}
\end{table}

\textbf{Targets.} We evaluate approaches against 6 guardrail systems listed in Table~\ref{tab:targets}, each comprising multiple scanners across detection categories (see Appendix Table \ref{tab:guardrail_detector_types}). We use each guardrail system default confidence score threshold and, where no default is provided, set the threshold to $0.50$. Five of the six evaluated guardrail systems are established open-source or commercial guardrail systems, while the sixth, \texttt{Ensemble Guardrail}, is assembled from three popular open-source HuggingFace models covering distinct detection categories: prompt injection and jailbreak \cite{roguesecurity2025sentinel}, toxicity \cite{hanu2020unbiasedtoxicroberta}, and gibberish \cite{jindal2021gibberishdetector}, to evaluate a custom-built guardrail system constructed of open-source models varying in architecture, training, and task.

\textbf{Bypass Candidates.} Candidates are selected from the \texttt{PIGuard} dataset, which are filtered to contain only malicious prompts across prompt injection and jailbreaks \cite{PIGuard}. We conducted an initial experiment to rank and select the top 20 prompts that were blocked by the highest number of scanners across our target guardrail systems, so the approaches are evaluated against the strictest bypass scenario. This results in 120 \textit{bypassing scenarios} across 6 guardrail systems, with 20 seed prompts per system, respectively.

\textbf{System Setup.} Experiments were conducted on a Linux Ubuntu 20.04 server with 128GB RAM, an Intel Xeon Gold 6336Y CPU, and two A100 GPU (80GB VRAM). Experiments scenarios were ran serially to minimize performance interference.

\textbf{Evaluation Metrics.} We evaluate our approach using three metrics. \textit{Attack Success Rate (ASR)}: The percentage of prompts that successfully evade all scanners in a guardrail system, \textit{Query Efficiency}: The average number of queries required to generate a successful bypass, and \textit{Wall-Clock Time}: The average time taken to generate a successful bypass.

\textbf{Bypass Similarity.} To evaluate how well the generated bypasses preserve the original seed prompts adversarial goal, we leverage two metrics: \textit{Surface Similarity}: A character-level sequence ratio between the original and bypass text, and \textit{Semantic Preservation}: The cosine similarity between sentence embeddings of the original and bypass text \cite{reimers-gurevych-2019-sentence}. Furthermore, to measure the \textit{adversarial utility} of the generated bypasses, we utilize an ablated \texttt{Qwen 3 8B model} selected as a control to observe whether the perturbation introduces divergence in model behaviour rather than the effectiveness of the bypass sample to overcome the model's safety alignment \cite{young2026comparativeanalysisllmabliteration}.To measure the utility, we use an LLM Judge, utilising \texttt{gpt-5.6-terra} to compare the ablated model's generated output when provided the original seed prompt and the generated bypass, scoring similarity from 1 to 5 \cite{llmjudges}.

\section{Results}
\label{branch:results}

\begin{figure*}[tbp]
    \centering
    \begin{subfigure}[t]{0.49\linewidth}
        \centering
        \includegraphics[width=\linewidth]{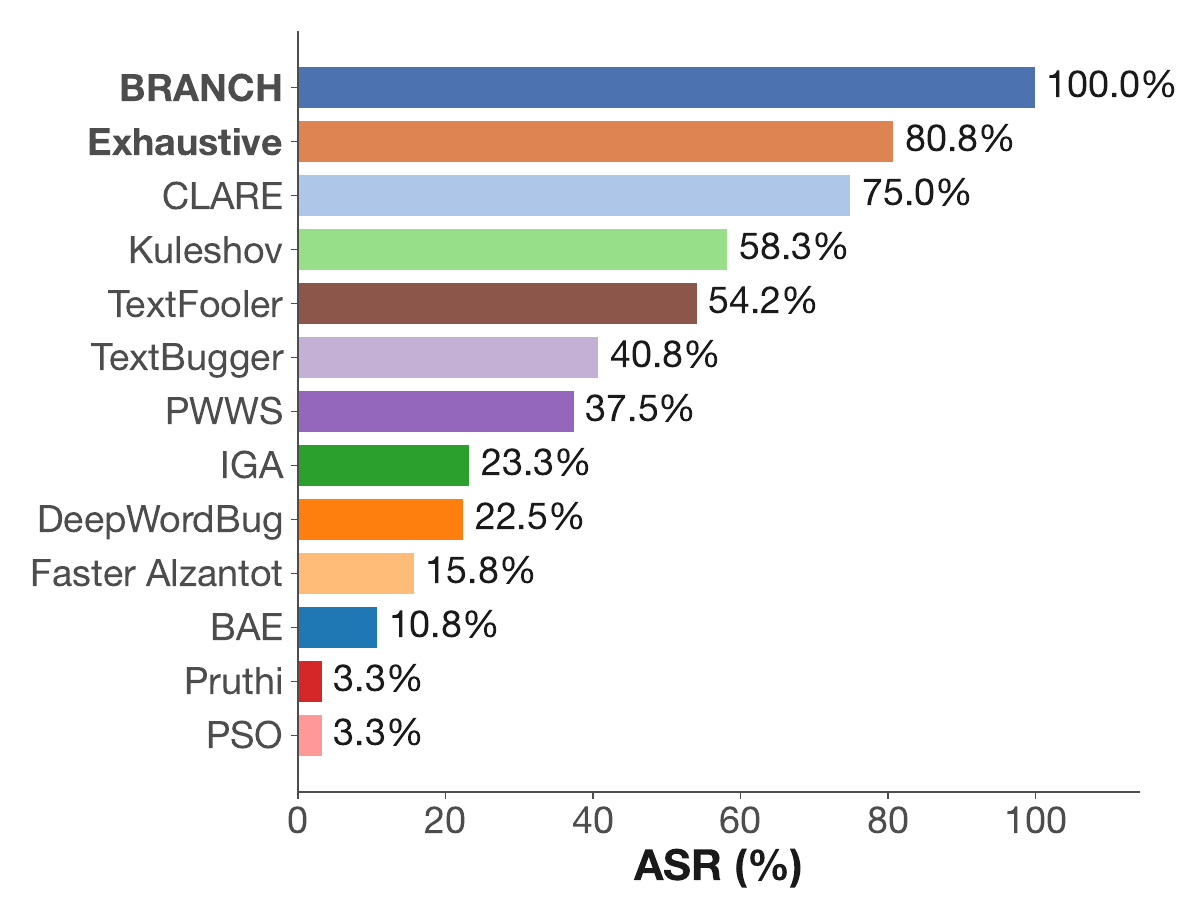}
        \caption{Bypass effectiveness.}
        \label{fig:success_rate_bars}
    \end{subfigure}
    \hfill
    \begin{subfigure}[t]{0.47\linewidth}
        \centering
        \includegraphics[width=\linewidth]{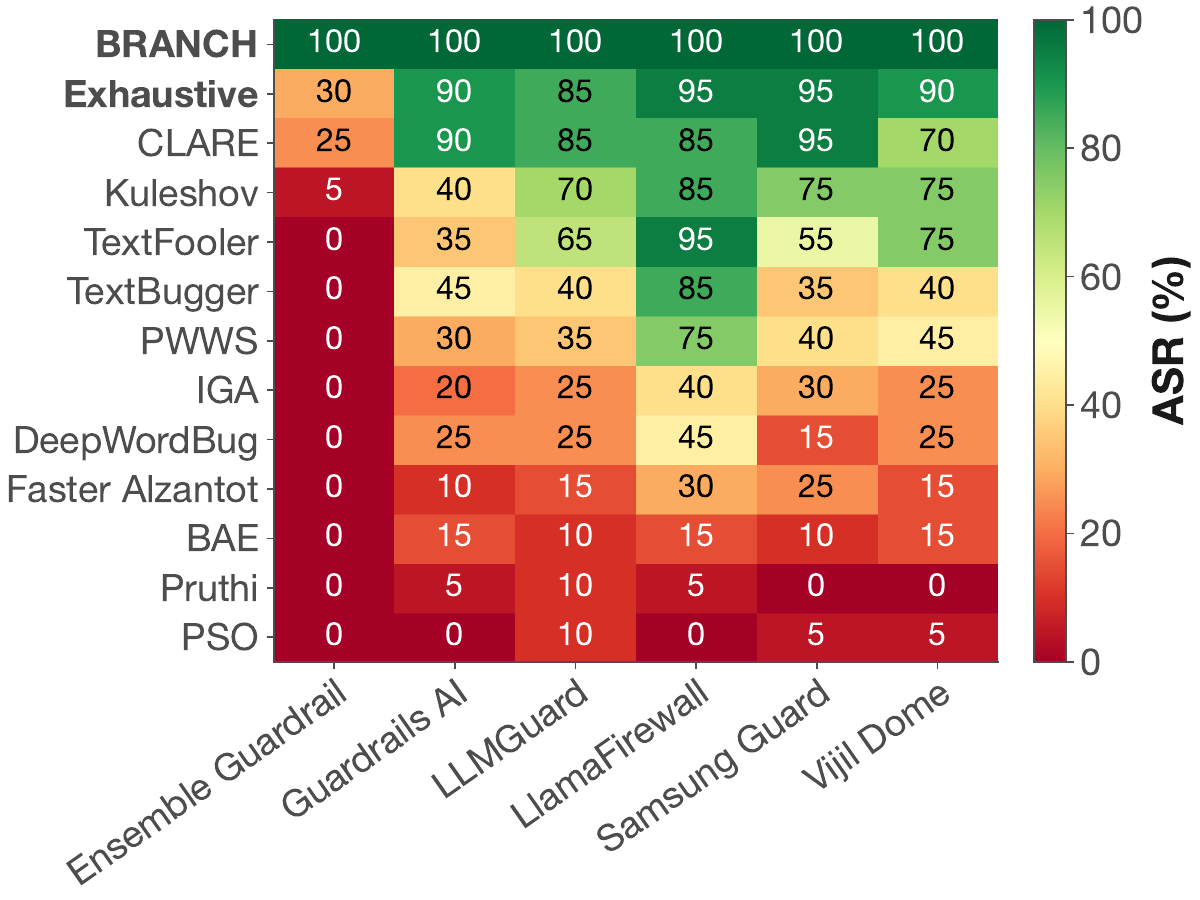}
        \caption{Heat map by guardrail target.}
        \label{fig:success_rate_heatmap}
    \end{subfigure}
    \caption{\textbf{Bypass Effectiveness.} Overall attack success rate of all methods against 6 guardrail systems.}
    \label{fig:success_rate}
\end{figure*}

\begin{figure*}
    \centering
    \begin{subfigure}[t]{0.49\linewidth}
        \centering
        \includegraphics[width=\linewidth]{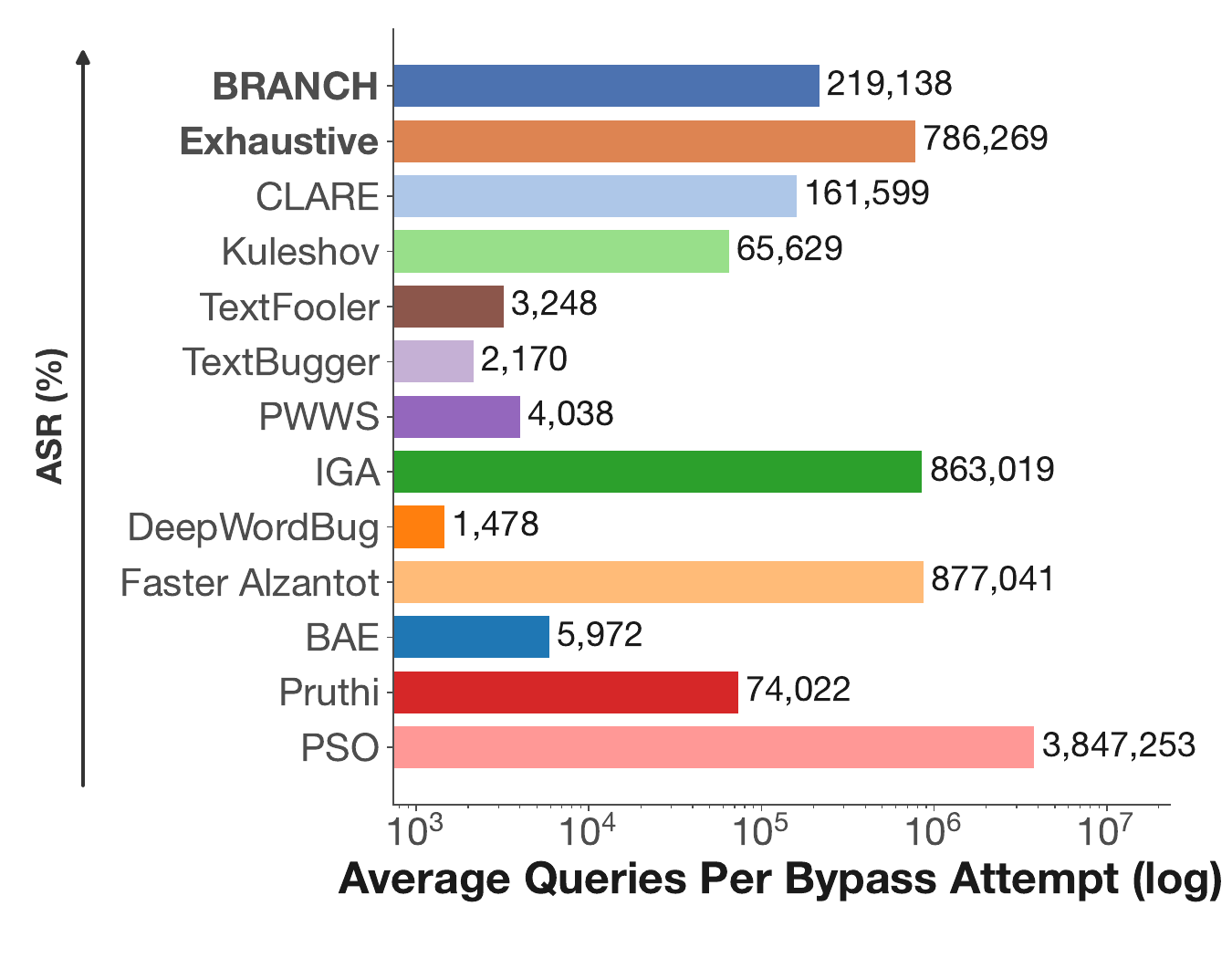}
        \caption{Query Efficiency.}
        \label{fig:efficiency_bars}
    \end{subfigure}
    \hfill
    \begin{subfigure}[t]{0.49\linewidth}
        \centering
        \includegraphics[width=\linewidth]{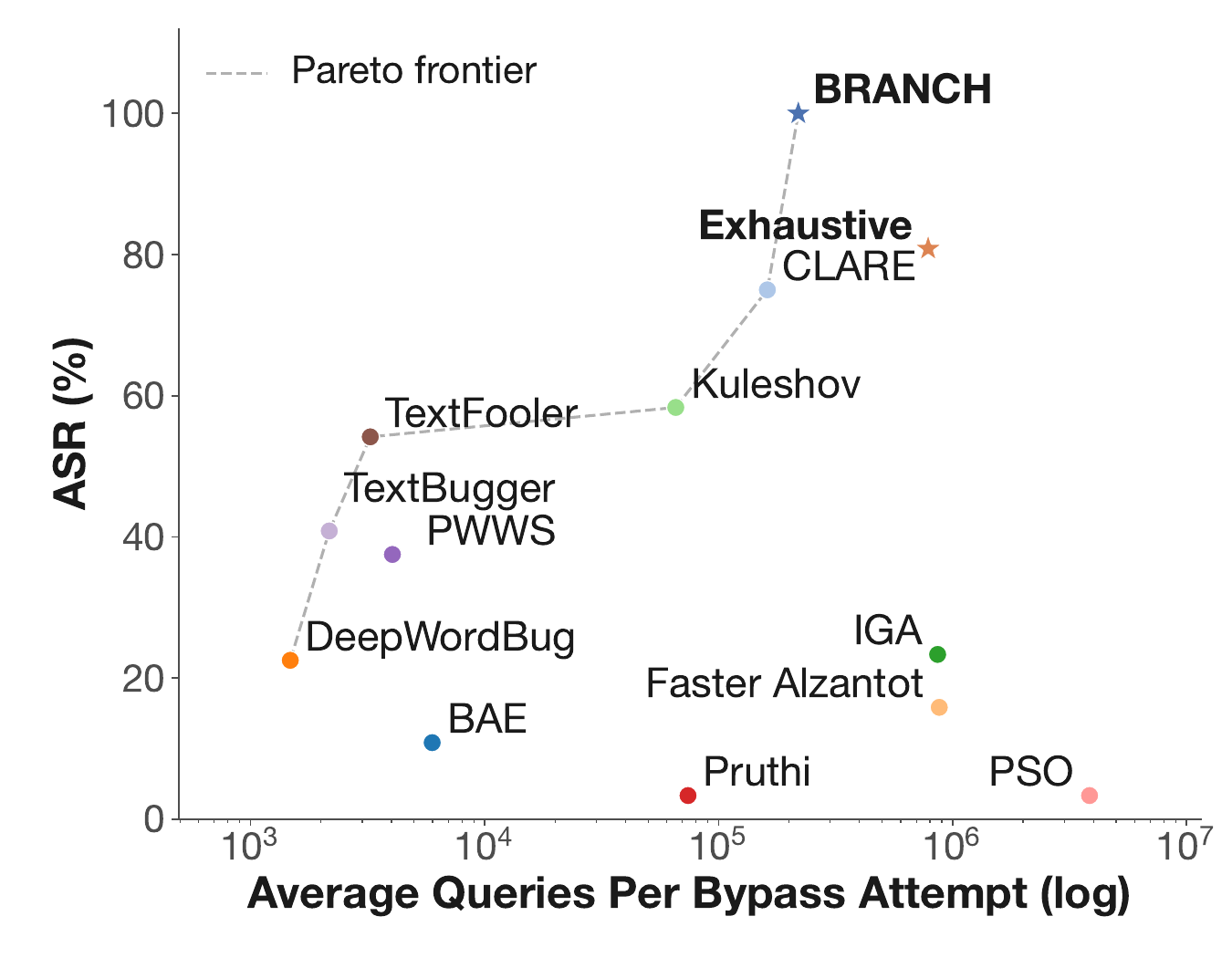}
        \caption{ASR vs Query efficiency.}
        \label{fig:efficiency_scatter}
    \end{subfigure}
    \caption{\textbf{Query Efficiency Across Methods.} Average queries per bypass attempt for BRANCH, Exhaustive, and the 11 individual techniques, across targeted guardrail system.}
    \label{fig:efficiency}
\end{figure*}

\subsection{Bypass Effectiveness}
\label{result:bypass_effectiveness}

\textbf{Attack Success Rate.} Figure~\ref{fig:success_rate_bars} shows the overall guardrail system bypass ASR across all approaches. BRANCH successfully generated a bypass in all 120 scenarios achieving 100\% ASR, compared to 80.8\% for Exhaustive. Focusing on individual techniques, CLARE was the most effective with an ASR of 75.0\%, followed by Kuleshov at 58.3\% and TextFooler at 54.2\%, while the weakest Pruthi and PSO attain 3.3\% each. Exhaustive reaches 80.8\%, 5.8\% higher than CLARE, indicating that the remaining 10 techniques therefore yielded limited additional coverage. Figure~\ref{fig:success_rate_heatmap} shows BRANCH reaches 100\% success across all 6 guardrail systems, while Exhaustive techniques average between 85\% and 95\% on five of these targets, with Ensemble Guardrail being the most robust against these methods at 30\%. We attribute the difficulty in bypassing Ensemble Guardrail due to a higher number of scanner activations compared to other targets (see Appendix Figure~\ref{fig:scanners_fired_combined}), where on average more than 1 scanner is activated initially leading to more overlapping boundaries. We observe that BRANCH's gains are additive, as across all bypass scenarios there is no singular bypass scenario whereby an Exhaustive technique succeeds that BRANCH fails against, bypassing 23 additional scenarios than Exhaustive, with CLARE accounting for the large majority of the coverage (see Appendix Figure~\ref{fig:coverage_head_to_head}). 

\begin{figure*}[t]
    \centering
    \includegraphics[width=\linewidth]{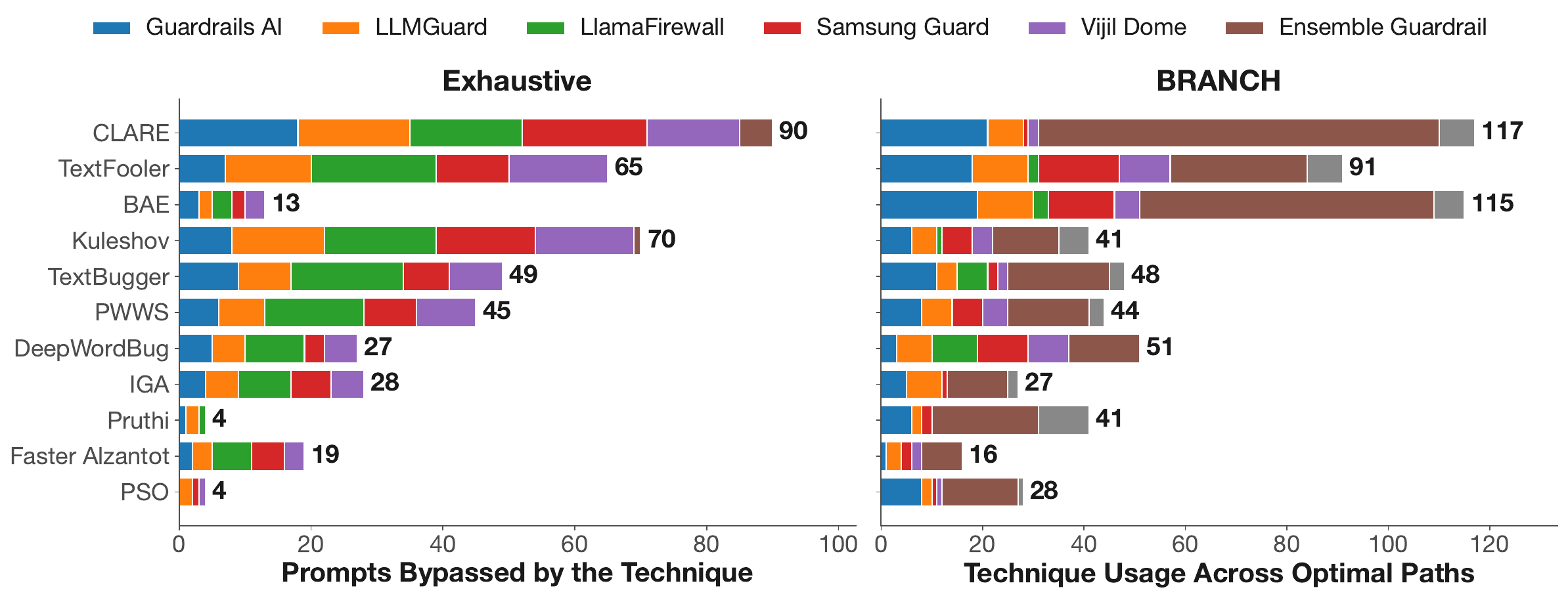}
    \caption{\textbf{Most Successful Perturbation Techniques.} The left panel shows the number of prompts bypassed by each Exhaustive technique alone. The right panel shows the number of times each technique is used along BRANCH's optimal paths.}
    \label{fig:technique_breakdown}
\end{figure*}

\textbf{Efficiency.} Figure~\ref{fig:efficiency_bars} shows that BRANCH spends 219,138 queries on average per bypass compared to 786,269 for Exhaustive, a 72.13\% reduction, or 3.58x fewer queries. Exhaustive's higher bypass cost is driven by scenarios it fails to bypass, running to exhaustion under an unbounded budget without ever finding a successful bypass. Even assuming identical budget between BRANCH and Exhaustive, this results in the average ASR of the latter dropping from 80.8\% to 75.0\% utilizing the most effective individual technique CLARE, and to 53.9\% if techniques are selected at random. Several individual techniques require fewer queries per bypass than BRANCH, including TextFooler and DeepWordBug, however attain substantially lower ASR by 45.8-89.2 percentage points. Figure~\ref{fig:efficiency_scatter} places every method on the ASR vs query-efficiency plane, where BRANCH lies on the top of the Pareto frontier. We observe that query cost is highest when targeting the Ensemble Guardrail across all bypassing techniques. Wall-clock time follows the same pattern, where BRANCH completes a bypass job in 2.9h compared to 13.1h for Exhaustive (4.51x speed-up). Among the individual techniques, the closest in ASR, CLARE, is slower at 5.8h, while faster techniques achieve lower ASRs (see Appendix Figure~\ref{fig:walltime}). While individual techniques such as CLARE can achieve high ASR, they do so only in specific bypass scenarios, meaning an adversary would be required to run multiple techniques in case one fails to achieve a bypass, therefore decreasing bypass efficiency. BRANCH is more efficient than Exhaustive, relaxing the requirement for an adversary to know which technique is best given a bypass scenario, achieving a higher ASR while consuming fewer queries and less wall-clock time.

\textbf{BRANCH Technique Composition.} We observe in Figure~\ref{fig:technique_breakdown} that BRANCH enabled less prominent bypassing techniques from Exhaustive to contribute meaningfully once chained together in BRANCH. For example, BAE and Pruthi-generated bypasses increase from 13 to 115 and 4 to 41, respectively within BRANCH's designation of the optimal bypass paths. In contrast, dominant techniques such as Kuleshov are used comparatively less (70 in Exhaustive, 41 in BRANCH). This indicates that techniques, especially those with lower ASR, contribute meaningfully to the final bypass within BRANCH, where when applied in isolation they struggle to bypass the target. We attribute this phenomenon to the specificity of bypass techniques. Where individual approaches typically utilize unique search and perturbation methodologies designed to exploit specific weaknesses within classification models. Chaining these techniques therefore enables them to share optimizations, creating bypasses which combine the advantages of multiple approaches. 

\begin{figure*}
    \centering
    \includegraphics[width=\linewidth]{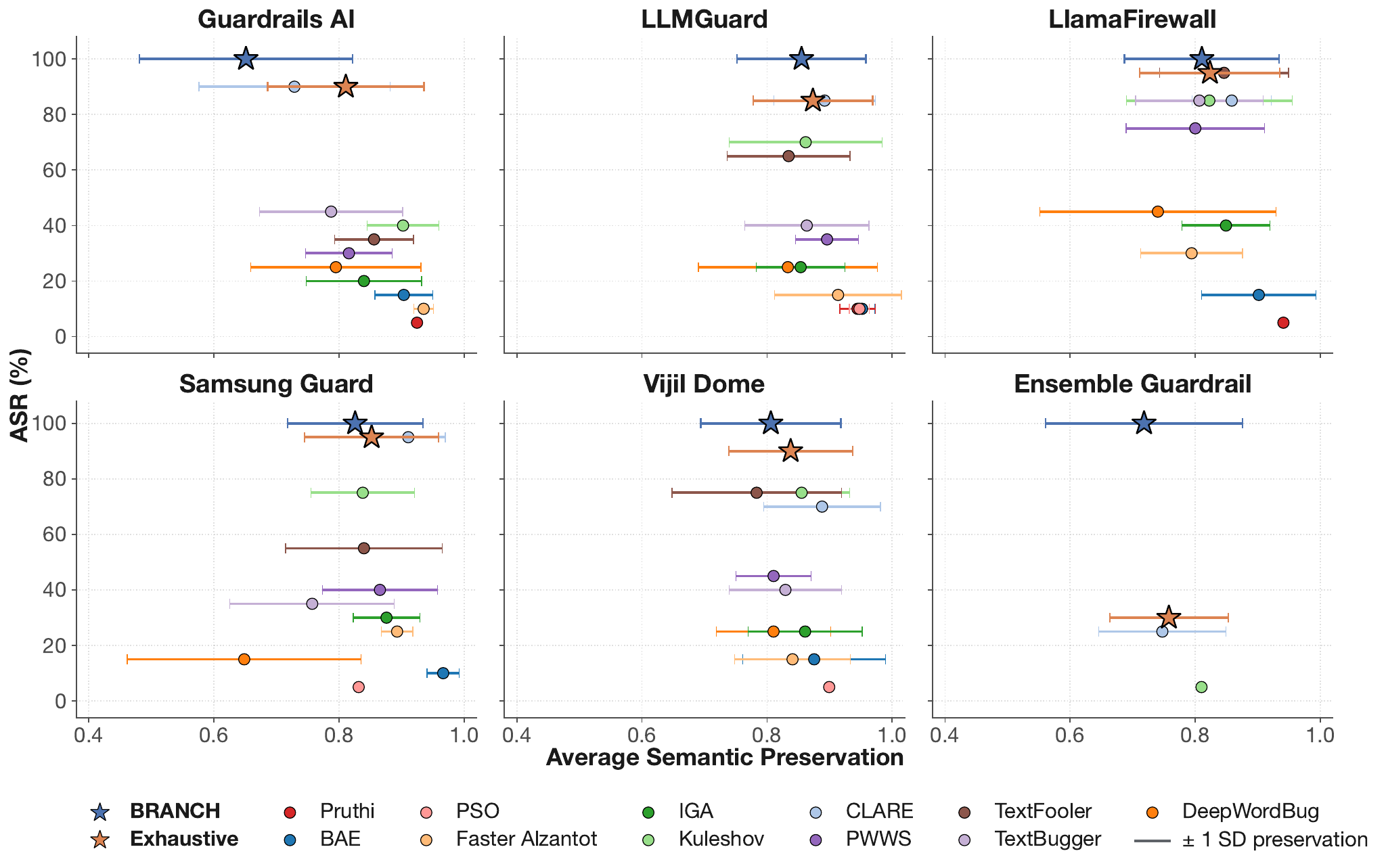}
    \caption{\textbf{Semantic Preservation.} Average semantic preservation of each technique's generated bypasses plotted against ASR. The stars mark BRANCH and Exhaustive, while dots represent individual techniques. Horizontal bars span $\pm 1$ standard deviation of semantic preservation across all generated bypasses.}
    \label{fig:similarity_overview}
\end{figure*}

\subsection{Generated Bypass Similarity}
\label{result:generated_bypass_similarity}

As stated in Section~\ref{branch:experiments}, beyond generating bypasses, it is important to determine whether the successful bypass still contains the same lexical meaning. We measured the original seed prompt to the generated bypass sample using two metrics: \textit{surface similarity} and \textit{semantic preservation}. We also measure the \textit{adversarial utility} of the bypass, observing if an ablated LLM interprets the bypass in the same way compared with the original seed prompt.


\newpage
\textbf{Surface Similarity and Semantic Preservation.} Figure~\ref{fig:similarity_overview} depicts the ASR and average semantic preservation of each approach. All 13 methods fall within a similar preservation deviation when evaluated across all targets. BRANCH achieves comparable average semantic preservation to Exhaustive across targets, such as 0.86 vs 0.87 for LLM Guard, 0.81 vs 0.82 for LlamaFirewall, 0.83 vs 0.85 for Vijil Dome, and 0.72 vs 0.76 for Ensemble, with the biggest difference against Guardrails AI at 0.65 vs 0.81 (Appendix Table~\ref{tab:similarity_semantic_values}). Surface similarity is where the two approaches greatly diverge, despite BRANCH preserving semantic content, the produced bypasses are modified more compared to Exhaustive and individual techniques, with BRANCH having an average surface similarity of 0.60 compared to 0.71 for Exhaustive (Appendix Figure~\ref{fig:similarity_per_target}). We attribute this to the side-effect of chaining perturbations together where the specific techniques such as IGA, and TextFooler appear to apply aggressively more modifications which can reduce surface similarity. Despite BRANCH applying more surface-level perturbations, the resulting bypass is still semantically similar to the seed prompt, and additionally comparable to established approaches. Therefore, not only does BRANCH achieve higher ASR compared to other bypass approaches, it does so while retaining the underlying meaning of the original prompt.

\begin{figure}[H]
    \centering
    \includegraphics[width=\linewidth]{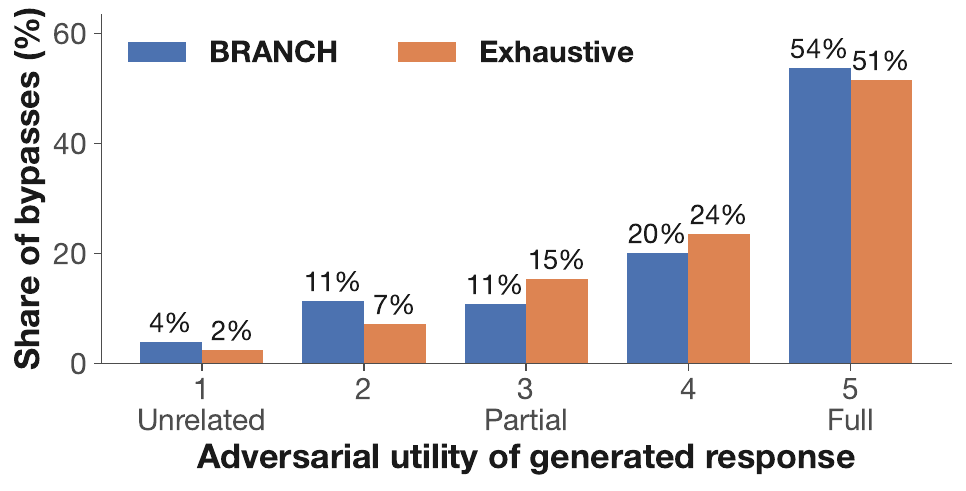}
    \caption{\textbf{Adversarial Utility.} Distribution of adversarial utility scores comparing the outputs generated from the original seed prompt and generated bypasses, evaluated by an LLM judge. Utility is scored from 1-5 (1 indicates no similarity, 5 indicates high similarity).}
    \label{fig:adversarial_utility}
\end{figure}

\textbf{BRANCH Depth Trade-off.} We further analyzed how surface similarity and semantic preservation decays as a function of tree depth (Appendix Figures~\ref{fig:depth_surface_similarity} and \ref{fig:depth_semantic_preservation}). Surface similarity decays sharply and continues to fall as depth increases across all 6 targets, from 1.0 to 0.23 by depth 21 on Guardrails AI, and to 0.14 by depth 13 on Samsung Guard. Semantic preservation decays far more gradually over the same depth range, typically plateauing after the first few layers, remaining between 0.85 and 0.88 across depths 2-10 on LLM Guard, before lowering to 0.72-0.77 in further depths. This gap between the two curves increases with depth, indicating that BRANCH accumulates larger surface-level perturbations as the depth of the tree grows while largely preserving the semantic content of the original request. As the semantic meaning is comparable to established approaches, this highlights that BRANCH is capable of exploring more sophisticated perturbation spaces compared to other techniques, modifying the underlying prompt more to achieve a bypass.  

\begin{figure*}[t]
    \centering
    \includegraphics[width=\linewidth]{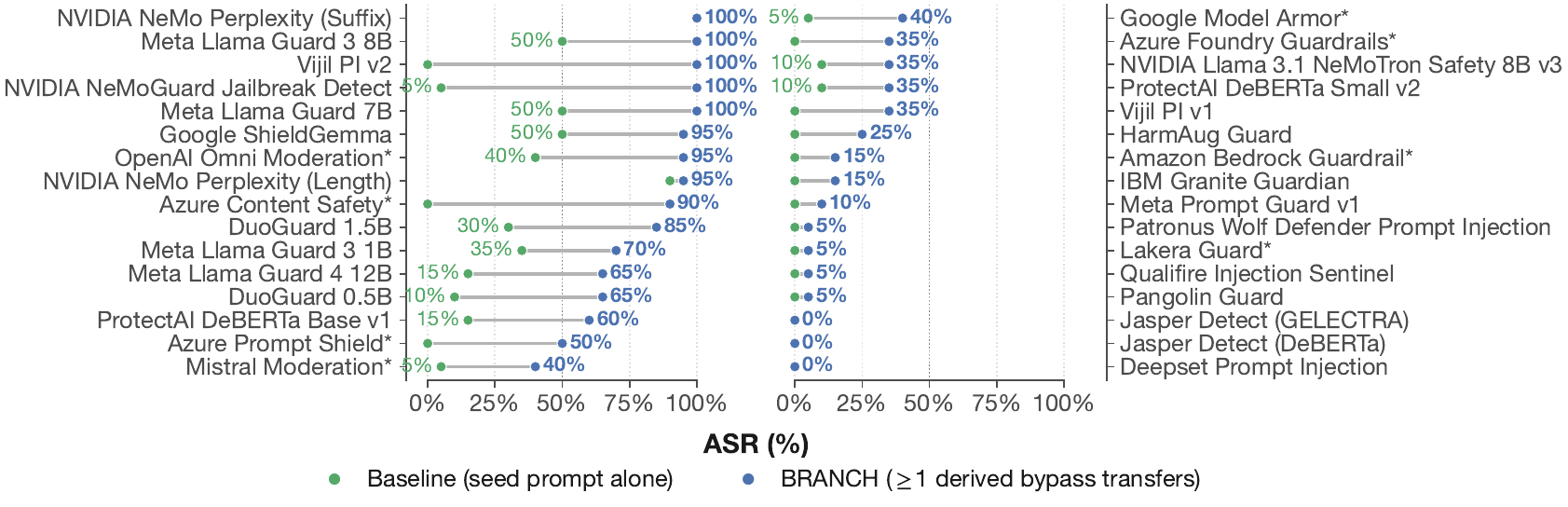}
    \caption{\textbf{Transferability to Unseen Guardrails.} Per-guardrail ASR of the 20 original seed prompts alone as a baseline against that of BRANCH's generated bypasses, for each of the 32 previously-unseen guardrails, ordered by transfer ASR. A seed prompt counts as transferring if at least one bypass derived from a given seed evades the guardrail. \textbf{*} = black-box commercial guardrails.}
    \label{fig:transferability_overview}
\end{figure*}

\textbf{Adversarial Utility.} We evaluated the adversarial utility of the generated bypasses against an ablated \texttt{Qwen 3 8B} model to determine whether the bypasses generate outputs which are similar to those from the original seed prompt. Figure~\ref{fig:adversarial_utility} shows that BRANCH generates comparable utility to the Exhaustive, with 85\% of BRANCH bypasses scoring at or above 3 compared to 90\% for Exhaustive, with 54\% of bypasses scoring 5 (indicating very high similarity) compared to 51\% for Exhaustive. These results highlight that the generated bypasses generate outputs against the ablated model which are similar to those generated using original seed prompts, indicating that intent is preserved through the bypassing process.

\subsection{Transferability}
\label{result:transferability}

To assess transferability of bypasses generated by BRANCH, we collect every successful bypass generated by BRANCH and evaluate them against a suite of 32 unseen white-box and black-box guardrail scanners, systems, and moderation LLMs. We measure transferability by assessing the number of bypasses derived from a seed prompt that evades an unseen guardrail.

\textbf{Transferability to Unseen Scanners.} As shown in  Figure~\ref{fig:transferability_overview}, we observe that BRANCH bypasses successfully transfers to 29 out of 32 unseen guardrails never queried during initial generation. We observe that the ASR per scanner ranges from 5\% on the most resistant scanners to 100\% on 5 scanners that are evaded on every seed prompt include (Meta Llama Guard 3 8B, NVIDIA NeMoGuard Jailbreak Detect, Vijil PI v2). Moreover, we note that generated bypasses transferring to unseen scanners appear to be exploiting shared weaknesses where specific targets are more likely to transfer to one another, rather than a generalized weakness across all scanners. For example, Guardrails AI appears to be the best target to generate bypasses that transfer to Meta Llama Guard 3 1B, and Meta Llama Guard 3 8B with 70\% and 77\% ASR, respectively, while Vijil Dome was the best target for Azure Foundry Guardrails and NVIDIA Llama 3.1 NeMoTron Safety Guard 8B v3 (see Appendix Figure~\ref{fig:transfer_heatmap_branch} for transfer ASR by target).

\textbf{Guardrail Transfer Effectiveness.} Utilizing BRANCH-generated prompts, ASR increased on average of 32.9 percentage points. We observe the largest ASR increases were NVIDIA NeMoGuard Jailbreak Detect and Vijil PI v2 (90\%), and Azure Content Safety (90\%). Scanners with moderate baselines behave similarly, including DuoGuard 0.5B (55\%) and OpenAI Omni Moderation from (55\%). 8 evaluated guardrails are commercial black-box offerings (no access to confidence scores) that exhibit an ASR increase within Lakera Guard (5\%), Amazon Bedrock Guardrail (15\%), Azure Foundry Guardrails (35\%), Google Model Armor and Mistral Moderation (35\%), Azure Prompt Shield (50\%), Azure Content Safety (90\%) and OpenAI Omni Moderation (55\%). 

These findings demonstrate that an attacker is capable of generating bypasses against a surrogate guardrail system and at no additional cost, transferring the bypasses to other unseen guardrails, including commercial black-box offerings.

\section{Limitations}

\textbf{Candidate Prompt Selection.} Our evaluation uses 20 seed prompts per guardrail system, totalling 120 bypass scenarios. These prompts were selected as those blocked by the most scanners across our targets (see Appendix Figure~\ref{fig:scanners_fired_combined}), ensuring experimentation was conducted within a strict multi-objective scenario. In doing so, full evaluation across the 120 bypass scenarios took a total of 1,923.8 hours, approximately 80 days. This cost was driven by the unbounded search used by Exhaustive and the individual techniques, which limited the number of bypass scenarios we could evaluate in favor or comparing to established work. While we believe these prompts sufficiently demonstrate the challenges faced by established techniques when targeting guardrail systems, and that BRANCH can address these limitations, future work can extend BRANCH to explore other candidate prompts sourced from other datasets. 

\textbf{Hyperparameter Optimization.} In our experiments, BRANCH uses a query budget of 5,000 per node greatly reduced from the unbounded budget utilized in Exhaustive. We believe their is opportunity to refine this further through optimization of the utilized query budget based upon parent nodes within the tree and the selected perturbation technique. Doing so could enable BRANCH to be more efficient given the context of the target and historical branches in the tree, enabling BRANCH to achieve bypasses earlier with fewer queries per node and terminating at a shallower depth.

\textbf{Threat Model.} BRANCH assumes an adversary can observe the per-scanner confidence scores and blocking decisions of the target guardrail system. This assumption is realistic when an adversary can access the guardrail system locally or if vendor APIs expose per-scanner outputs, typically offered by commercial guardrails to enable developers to configure detection thresholds, however would be less effective in strict black-box production scenarios where confidence scores and guardrail activation is hidden \cite{hackett2026refusaldeterminingguardrailactivation}. While this limitation is focused upon the generation of the bypass samples, our transferability study highlights that even in a strict black-box threat models, an adversary can generate bypasses offline which successfully transfer to an unseen target without requiring to perform optimization (Section~\ref{result:transferability}).

\textbf{Bypass Efficacy.} Our study focuses primarily on the task of bypassing guardrails. While we interoperate various metrics to measure adversarial utility, such as semantic preservation, surface similarity, and utilizing an ablated model to observe if the intent of the bypass prompt matches that of the original, we do not conduct a thorough experiment on the efficacy of the bypassed prompt against production LLMs. Therefore, we believe there exists further work to assess the utility of the bypassed prompts and the effect BRANCH and various techniques have on the overall ASRs of the prompts.

\section{Related Work}
\label{related_work}

Adversarial ML evasion techniques have been used to successfully bypass text-classification models \cite{pruthi, textbugger, deepwordbug, kuleshov, iga} and more recently, guardrails.  However, their respective evaluation focus isolated single-scanner guardrails rather than multi-scanner guardrail systems \cite{hackettbypassing, guardrail-mismatch}. For example, early techniques such as TextFooler \cite{textfooler}, BERT-Attack \cite{bertattack}, and PWWS \cite{pwws} demonstrated their effectiveness against sentiment analysis and classification models, which were later applied to widely used classification models in guardrails \cite{hackettbypassing}. More recent guardrail bypassing approaches exploit different weaknesses, such as hiding content outside the input window of the guardrail \cite{zhou2026promptoverflowguardrailinspects}, or exploiting representation asymmetry between guardrail classifiers and the LLM, where minor character perturbations that confuse the guardrail remain interpretable to the target model \cite{guardrail-mismatch}. All of these studies are evaluated against multiple single-scanner guardrails,  and do not consider scenarios whereby multiple scanners are deployed within a guardrail system that require bypasses to evade all scanners in order to succeed.

\section{Conclusion}
\label{conclusion}

In this paper we have proposed BRANCH -- a methodology designed to bypass multi-scanner guardrail systems. Our work identifies and targets the gap in current evasion attack and guardrail research where existing techniques struggle to converge on adversarial prompts capable of bypassing guardrail systems containing multiple scanners. Across 6 target guardrail systems and 120 bypass scenarios, BRANCH is capable of attaining 100\% bypass success rate while consuming 72\% fewer queries and 4.5x less wall-clock time per bypass. Importantly, we demonstrate that generated bypasses transfer reasonably well to a large number of unseen guardrail systems, including commercial offerings. We hope our findings will help the research community to create more advanced evasion attack techniques applicable to more sophisticated AI system defenses, as well as help practitioners evaluate their guardrail system effectiveness.

\section{Acknowledgments}

We would like to thank members of the Lancaster University AI Security Lab and Mindgard research team for their comments and feedback. This work was supported by the UK Engineering Physical Science and Research Council (EPSRC) and Mindgard.

\bibliography{paper_references}

@misc{dong2024safeguardinglargelanguagemodels,
      title={Safeguarding Large Language Models: A Survey}, 
      author={Yi Dong and Ronghui Mu and Yanghao Zhang and Siqi Sun and Tianle Zhang and Changshun Wu and Gaojie Jin and Yi Qi and Jinwei Hu and Jie Meng and Saddek Bensalem and Xiaowei Huang},
      year={2024},
      eprint={2406.02622},
      archivePrefix={arXiv},
      primaryClass={cs.CR},
      url={https://arxiv.org/abs/2406.02622}, 
}

@inproceedings{morris2020textattack,
  title={TextAttack: A Framework for Adversarial Attacks, Data Augmentation, and Adversarial Training in NLP},
  author={Morris, John and Lifland, Eli and Yoo, Jin Yong and Grigsby, Jake and Jin, Di and Qi, Yanjun},
  booktitle={Proceedings of the 2020 Conference on Empirical Methods in Natural Language Processing: System Demonstrations},
  pages={119--126},
  year={2020}
}

@book{miettinen1999nonlinear,
  title={Nonlinear Multiobjective Optimization},
  author={Miettinen, Kaisa},
  year={1999},
  publisher={Springer},
  address={Boston, MA},
  series={International Series in Operations Research \& Management Science},
  volume={12}
}

@misc{dubey2024llama3herdmodels,
      title={The Llama 3 Herd of Models},
      author={Abhimanyu Dubey and et al.},
      year={2024},
      eprint={2407.21783},
      archivePrefix={arXiv},
      primaryClass={cs.AI},
      url={https://arxiv.org/abs/2407.21783}, 
}

@misc{nemoguardrails,
      title={NeMo Guardrails: A Toolkit for Controllable and Safe LLM Applications with Programmable Rails}, 
      author={Traian Rebedea and Razvan Dinu and Makesh Sreedhar and Christopher Parisien and Jonathan Cohen},
      year={2023},
      eprint={2310.10501},
      archivePrefix={arXiv},
      primaryClass={cs.CL},
      url={https://arxiv.org/abs/2310.10501}, 
}

@misc{roguesecurity2025sentinel,
  author = {{Rogue Security}},
  title = {Prompt Injection \& Jailbreak Sentinel v2},
  year = {2025},
  publisher = {HuggingFace},
  url = {https://huggingface.co/rogue-security/prompt-injection-jailbreak-sentinel-v2},
}

@misc{hanu2020unbiasedtoxicroberta,
  author = {Hanu, Laura and {Unitary team}},
  title = {Unbiased Toxic RoBERTa},
  year = {2020},
  publisher = {HuggingFace},
  url = {https://huggingface.co/unitary/unbiased-toxic-roberta},
}

@misc{jindal2021gibberishdetector,
  author = {Jindal, Madhur},
  title = {AutoNLP Gibberish Detector},
  year = {2021},
  publisher = {HuggingFace},
  url = {https://huggingface.co/madhurjindal/autonlp-Gibberish-Detector-492513457},
}

@misc{bertattack,
      title={BERT-ATTACK: Adversarial Attack Against BERT Using BERT}, 
      author={Linyang Li and Ruotian Ma and Qipeng Guo and Xiangyang Xue and Xipeng Qiu},
      year={2020},
      eprint={2004.09984},
      archivePrefix={arXiv},
      primaryClass={cs.CL},
      url={https://arxiv.org/abs/2004.09984}, 
}

@inproceedings{bae,
    title = "{BAE}: {BERT}-based Adversarial Examples for Text Classification",
    author = "Garg, Siddhant  and
      Ramakrishnan, Goutham",
    editor = "Webber, Bonnie  and
      Cohn, Trevor  and
      He, Yulan  and
      Liu, Yang",
    booktitle = "Proceedings of the 2020 Conference on Empirical Methods in Natural Language Processing (EMNLP)",
    month = nov,
    year = "2020",
    address = "Online",
    publisher = "Association for Computational Linguistics",
    url = "https://aclanthology.org/2020.emnlp-main.498",
    doi = "10.18653/v1/2020.emnlp-main.498",
    pages = "6174--6181",
}

@misc{deepwordbug,
      title={Black-box Generation of Adversarial Text Sequences to Evade Deep Learning Classifiers}, 
      author={Ji Gao and Jack Lanchantin and Mary Lou Soffa and Yanjun Qi},
      year={2018},
      eprint={1801.04354},
      archivePrefix={arXiv},
      primaryClass={cs.CL},
      url={https://arxiv.org/abs/1801.04354}, 
}

@inproceedings{pso,
    title = "Word-level Textual Adversarial Attacking as Combinatorial Optimization",
    author = "Zang, Yuan  and
      Qi, Fanchao  and
      Yang, Chenghao  and
      Liu, Zhiyuan  and
      Zhang, Meng  and
      Liu, Qun  and
      Sun, Maosong",
    editor = "Jurafsky, Dan  and
      Chai, Joyce  and
      Schluter, Natalie  and
      Tetreault, Joel",
    booktitle = "Proceedings of the 58th Annual Meeting of the Association for Computational Linguistics",
    month = jul,
    year = "2020",
    address = "Online",
    publisher = "Association for Computational Linguistics",
    url = "https://aclanthology.org/2020.acl-main.540/",
    doi = "10.18653/v1/2020.acl-main.540",
    pages = "6066--6080"
}

@inproceedings{clare,
    title = "Contextualized Perturbation for Textual Adversarial Attack",
    author = "Li, Dianqi  and
      Zhang, Yizhe  and
      Peng, Hao  and
      Chen, Liqun  and
      Brockett, Chris  and
      Sun, Ming-Ting  and
      Dolan, Bill",
    editor = "Toutanova, Kristina  and
      Rumshisky, Anna  and
      Zettlemoyer, Luke  and
      Hakkani-Tur, Dilek  and
      Beltagy, Iz  and
      Bethard, Steven  and
      Cotterell, Ryan  and
      Chakraborty, Tanmoy  and
      Zhou, Yichao",
    booktitle = "Proceedings of the 2021 Conference of the North American Chapter of the Association for Computational Linguistics: Human Language Technologies",
    month = jun,
    year = "2021",
    address = "Online",
    publisher = "Association for Computational Linguistics",
    url = "https://aclanthology.org/2021.naacl-main.400/",
    doi = "10.18653/v1/2021.naacl-main.400",
    pages = "5053--5069"
}

@misc{iga,
      title={Natural Language Adversarial Defense through Synonym Encoding}, 
      author={Xiaosen Wang and Hao Jin and Yichen Yang and Kun He},
      year={2021},
      eprint={1909.06723},
      archivePrefix={arXiv},
      primaryClass={cs.CL},
      url={https://arxiv.org/abs/1909.06723}, 
}

@inproceedings{fasteralzantot,
    title = "Certified Robustness to Adversarial Word Substitutions",
    author = {Jia, Robin  and
      Raghunathan, Aditi  and
      G{\"o}ksel, Kerem  and
      Liang, Percy},
    editor = "Inui, Kentaro  and
      Jiang, Jing  and
      Ng, Vincent  and
      Wan, Xiaojun",
    booktitle = "Proceedings of the 2019 Conference on Empirical Methods in Natural Language Processing and the 9th International Joint Conference on Natural Language Processing (EMNLP-IJCNLP)",
    month = nov,
    year = "2019",
    address = "Hong Kong, China",
    publisher = "Association for Computational Linguistics",
    url = "https://aclanthology.org/D19-1423/",
    doi = "10.18653/v1/D19-1423",
    pages = "4129--4142"
}

@misc{kuleshov,
title={Adversarial Examples for Natural Language Classification Problems},
author={Volodymyr Kuleshov and Shantanu Thakoor and Tingfung Lau and Stefano Ermon},
year={2018},
url={https://openreview.net/forum?id=r1QZ3zbAZ},
}

@misc{alzantot,
      title={Generating Natural Language Adversarial Examples},
      author={Moustafa Alzantot and Yash Sharma and Ahmed Elgohary and Bo-Jhang Ho and Mani Srivastava and Kai-Wei Chang},
      year={2018},
      eprint={1804.07998},
      archivePrefix={arXiv},
      primaryClass={cs.CL},
      url={https://arxiv.org/abs/1804.07998}, 
}

@misc{textfooler,
      title={Is BERT Really Robust? A Strong Baseline for Natural Language Attack on Text Classification and Entailment}, 
      author={Di Jin and Zhijing Jin and Joey Tianyi Zhou and Peter Szolovits},
      year={2020},
      eprint={1907.11932},
      archivePrefix={arXiv},
      primaryClass={cs.CL},
      url={https://arxiv.org/abs/1907.11932}, 
}

@inproceedings{pwws,
    title = "Generating Natural Language Adversarial Examples through Probability Weighted Word Saliency",
    author = "Ren, Shuhuai  and
      Deng, Yihe  and
      He, Kun  and
      Che, Wanxiang",
    editor = "Korhonen, Anna  and
      Traum, David  and
      M{\`a}rquez, Llu{\'\i}s",
    booktitle = "Proceedings of the 57th Annual Meeting of the Association for Computational Linguistics",
    month = jul,
    year = "2019",
    address = "Florence, Italy",
    publisher = "Association for Computational Linguistics",
    url = "https://aclanthology.org/P19-1103",
    doi = "10.18653/v1/P19-1103",
    pages = "1085--1097",
}

@misc{pruthi,
      title={Combating Adversarial Misspellings with Robust Word Recognition}, 
      author={Danish Pruthi and Bhuwan Dhingra and Zachary C. Lipton},
      year={2019},
      eprint={1905.11268},
      archivePrefix={arXiv},
      primaryClass={cs.CL},
      url={https://arxiv.org/abs/1905.11268}, 
}

@inproceedings{textbugger, series={NDSS 2019},
   title={TextBugger: Generating Adversarial Text Against Real-world Applications},
   url={http://dx.doi.org/10.14722/ndss.2019.23138},
   DOI={10.14722/ndss.2019.23138},
   booktitle={Proceedings 2019 Network and Distributed System Security Symposium},
   publisher={Internet Society},
   author={Li, Jinfeng and Ji, Shouling and Du, Tianyu and Li, Bo and Wang, Ting},
   year={2019},
   collection={NDSS 2019} }

@misc{microsoft2024contentfilter,
  title = {Content Filtering in Azure OpenAI Service},
  author = {{Microsoft Corporation}},
  year = {2024},
  url = {https://learn.microsoft.com/en-us/azure/ai-services/openai/concepts/content-filter?tabs=warning%2Cpython-new},
  note = {Accessed: 2024-08-06}
}

@misc{demontis2019adversarialattackstransferexplaining,
      title={Why Do Adversarial Attacks Transfer? Explaining Transferability of Evasion and Poisoning Attacks}, 
      author={Ambra Demontis and Marco Melis and Maura Pintor and Matthew Jagielski and Battista Biggio and Alina Oprea and Cristina Nita-Rotaru and Fabio Roli},
      year={2019},
      eprint={1809.02861},
      archivePrefix={arXiv},
      primaryClass={cs.LG},
      url={https://arxiv.org/abs/1809.02861}, 
}

@inproceedings{llmjudges,
author = {Zheng, Lianmin and Chiang, Wei-Lin and Sheng, Ying and Zhuang, Siyuan and Wu, Zhanghao and Zhuang, Yonghao and Lin, Zi and Li, Zhuohan and Li, Dacheng and Xing, Eric P. and Zhang, Hao and Gonzalez, Joseph E. and Stoica, Ion},
title = {Judging LLM-as-a-judge with MT-bench and Chatbot Arena},
year = {2024},
publisher = {Curran Associates Inc.},
address = {Red Hook, NY, USA},
booktitle = {Proceedings of the 37th International Conference on Neural Information Processing Systems},
articleno = {2020},
numpages = {29},
location = {New Orleans, LA, USA},
series = {NIPS '23}
}

@misc{llmguard2025,
  author       = {{Protect AI}},
  title        = {LLM Guard: The Security Toolkit for LLM Interactions},
  howpublished = {\url{https://github.com/protectai/llm-guard}},
  url = {https://github.com/protectai/llm-guard},
  note         = {Accessed: 2026-07-04},
  year         = {2025}
}

@misc{vijildome2024,
  author       = {{Vijil AI}},
  title        = {Vijil Dome},
  howpublished = {\url{https://github.com/vijilAI/vijil-dome}},
  url = {https://github.com/vijilAI/vijil-dome},
  note         = {Accessed: 2026-07-04},
  year         = {2024}
}

@misc{guardrailsai,
  author       = {{Guardrails AI}},
  title        = {Guardrails AI},
  howpublished = {\url{https://guardrailsai.com/}},
  url = {https://guardrailsai.com/},
  note         = {Accessed: 2026-07-04},
  year         = {2026}
}

@InProceedings{evasion_attacks,
author="Biggio, Battista
and Corona, Igino
and Maiorca, Davide
and Nelson, Blaine
and {\v{S}}rndi{\'{c}}, Nedim
and Laskov, Pavel
and Giacinto, Giorgio
and Roli, Fabio",
editor="Blockeel, Hendrik
and Kersting, Kristian
and Nijssen, Siegfried
and {\v{Z}}elezn{\'y}, Filip",
title="Evasion Attacks against Machine Learning at Test Time",
booktitle="Machine Learning and Knowledge Discovery in Databases",
year="2013",
publisher="Springer Berlin Heidelberg",
address="Berlin, Heidelberg",
pages="387--402",
isbn="978-3-642-40994-3"
}

@article{PIGuard,
  title={PIGuard: Prompt Injection Guardrail via Mitigating Overdefense for Free},
  author={Hao Li and 
        Xiaogeng Liu and 
        Ning Zhang and 
        Chaowei Xiao},
  journal = {ACL},
  year={2025}
}

@misc{zhou2026promptoverflowguardrailinspects,
      title={Prompt Overflow: What the Guardrail Inspects Is Not What the Model Infers}, 
      author={Yuanbo Zhou and Changjia Zhu and Junyu Wang and Xu He and Yan Zhai and Kun Sun and Mingkui Wei and Junjie Xiong},
      year={2026},
      eprint={2605.23196},
      archivePrefix={arXiv},
      primaryClass={cs.CR},
      url={https://arxiv.org/abs/2605.23196}, 
}

@inproceedings{guardrail-mismatch,
author = {Wang, Junyi and Zhu, Zhibin and Liu, Chuanyi},
title = {The Asymmetric Vulnerability: Bypassing LLM Defenses via Guardrail-Model Mismatch},
year = {2026},
isbn = {9798400723070},
publisher = {Association for Computing Machinery},
address = {New York, NY, USA},
url = {https://doi.org/10.1145/3774904.3792438},
doi = {10.1145/3774904.3792438},
booktitle = {Proceedings of the ACM Web Conference 2026},
pages = {3078–3089},
numpages = {12},
location = {United Arab Emirates},
series = {WWW '26}
}

@misc{llamafirewall,
      title={LlamaFirewall: An open source guardrail system for building secure AI agents}, 
      author={Sahana Chennabasappa and Cyrus Nikolaidis and Daniel Song and David Molnar and Stephanie Ding and Shengye Wan and Spencer Whitman and Lauren Deason and Nicholas Doucette and Abraham Montilla and Alekhya Gampa and Beto de Paola and Dominik Gabi and James Crnkovich and Jean-Christophe Testud and Kat He and Rashnil Chaturvedi and Wu Zhou and Joshua Saxe},
      year={2025},
      eprint={2505.03574},
      archivePrefix={arXiv},
      primaryClass={cs.CR},
      url={https://arxiv.org/abs/2505.03574}, 
}

@misc{SGuard-v1,
      title={SGuard-v1: Safety Guardrail for Large Language Models}, 
      author={JoonHo Lee and HyeonMin Cho and Jaewoong Yun and Hyunjae Lee and JunKyu Lee and Juree Seok},
      year={2025},
      eprint={2511.12497},
      archivePrefix={arXiv},
      primaryClass={cs.CL}
}

@misc{rossi2024earlycategorizationpromptinjection,
      title={An Early Categorization of Prompt Injection Attacks on Large Language Models}, 
      author={Sippo Rossi and Alisia Marianne Michel and Raghava Rao Mukkamala and Jason Bennett Thatcher},
      year={2024},
      eprint={2402.00898},
      archivePrefix={arXiv},
      primaryClass={cs.CR},
      url={https://arxiv.org/abs/2402.00898}, 
}

@article{GENG2026,
title = {Prompt Injection Attacks on Large Language Models: A Survey of Attack Methods, Root Causes, and Defense Strategies},
journal = {Computers, Materials and Continua},
volume = {87},
number = {1},
year = {2026},
issn = {1546-2218},
doi = {https://doi.org/10.32604/cmc.2025.074081},
url = {https://www.sciencedirect.com/science/article/pii/S1546221826001384},
author = {Tongcheng Geng and Zhiyuan Xu and Yubin Qu and W. Eric Wong}
}

@article{shethiya2023llm,
  title={LLM-Powered Architectures: Designing the Next Generation of Intelligent Software Systems},
  author={Shethiya, Aditya S},
  journal={Academia Nexus Journal},
  volume={2},
  number={1},
  year={2023}
}

@article{ayyamperumal2024current,
  title={Current state of llm risks and ai guardrails},
  author={Ayyamperumal, Suriya Ganesh and Ge, Limin},
  journal={arXiv preprint arXiv:2406.12934},
  year={2024}
}

@article{wu2024new,
  title={A new era in llm security: Exploring security concerns in real-world llm-based systems},
  author={Wu, Fangzhou and Zhang, Ning and Jha, Somesh and McDaniel, Patrick and Xiao, Chaowei},
  journal={arXiv preprint arXiv:2402.18649},
  year={2024}
}

@inproceedings{bird-loper-2004-nltk,
    title = "{NLTK}: The Natural Language Toolkit",
    author = "Bird, Steven  and
      Loper, Edward",
    booktitle = "Proceedings of the {ACL} Interactive Poster and Demonstration Sessions",
    month = jul,
    year = "2004",
    address = "Barcelona, Spain",
    publisher = "Association for Computational Linguistics",
    url = "https://aclanthology.org/P04-3031/",
    pages = "214--217"
}

@misc{hackett2026refusaldeterminingguardrailactivation,
      title={Behind the Refusal: Determining Guardrail Activation via Behavioral Monitoring}, 
      author={William Hackett and Peter Garraghan},
      year={2026},
      eprint={2607.02121},
      archivePrefix={arXiv},
      primaryClass={cs.CR},
      url={https://arxiv.org/abs/2607.02121}, 
}

@Article{electronics14132717,
AUTHOR = {Nastoska, Aleksandra and Jancheska, Bojana and Rizinski, Maryan and Trajanov, Dimitar},
TITLE = {Evaluating Trustworthiness in AI: Risks, Metrics, and Applications Across Industries},
JOURNAL = {Electronics},
VOLUME = {14},
YEAR = {2025},
NUMBER = {13},
ARTICLE-NUMBER = {2717},
URL = {https://www.mdpi.com/2079-9292/14/13/2717},
ISSN = {2079-9292},
DOI = {10.3390/electronics14132717}
}

@article{bassani2025guardrail,
  title={On Guardrail Models’ Robustness to Mutations and Adversarial Attacks},
  author={Bassani, Elias and Sanchez, Ignacio and others},
  journal={Findings of the Association for Computational Linguistics: EMNLP},
  volume={2025},
  year={2025}
}

@inproceedings{reimers-gurevych-2019-sentence,
    title = "Sentence-{BERT}: Sentence Embeddings using {S}iamese {BERT}-Networks",
    author = "Reimers, Nils  and
      Gurevych, Iryna",
    editor = "Inui, Kentaro  and
      Jiang, Jing  and
      Ng, Vincent  and
      Wan, Xiaojun",
    booktitle = "Proceedings of the 2019 Conference on Empirical Methods in Natural Language Processing and the 9th International Joint Conference on Natural Language Processing (EMNLP-IJCNLP)",
    month = nov,
    year = "2019",
    address = "Hong Kong, China",
    publisher = "Association for Computational Linguistics",
    url = "https://aclanthology.org/D19-1410/",
    doi = "10.18653/v1/D19-1410",
    pages = "3982--3992"
}

@misc{markov2023holisticapproachundesiredcontent,
      title={A Holistic Approach to Undesired Content Detection in the Real World}, 
      author={Todor Markov and Chong Zhang and Sandhini Agarwal and Tyna Eloundou and Teddy Lee and Steven Adler and Angela Jiang and Lilian Weng},
      year={2023},
      eprint={2208.03274},
      archivePrefix={arXiv},
      primaryClass={cs.CL},
      url={https://arxiv.org/abs/2208.03274}, 
}

@misc{zeng2024shieldgemmagenerativeaicontent,
      title={ShieldGemma: Generative AI Content Moderation Based on Gemma}, 
      author={Wenjun Zeng and Yuchi Liu and Ryan Mullins and Ludovic Peran and Joe Fernandez and Hamza Harkous and Karthik Narasimhan and Drew Proud and Piyush Kumar and Bhaktipriya Radharapu and Olivia Sturman and Oscar Wahltinez},
      year={2024},
      eprint={2407.21772},
      archivePrefix={arXiv},
      primaryClass={cs.CL},
      url={https://arxiv.org/abs/2407.21772}, 
}

@misc{young2026comparativeanalysisllmabliteration,
      title={Comparative Analysis of LLM Abliteration Methods: A Cross-Architecture Evaluation}, 
      author={Richard J. Young},
      year={2026},
      eprint={2512.13655},
      archivePrefix={arXiv},
      primaryClass={cs.CL},
      url={https://arxiv.org/abs/2512.13655}, 
}

@inproceedings{hackettbypassing,
    title = "Bypassing {LLM} Guardrails: An Empirical Analysis of Evasion Attacks against Prompt Injection and Jailbreak Detection Systems",
    author = "Hackett, William  and
      Birch, Lewis  and
      Trawicki, Stefan  and
      Suri, Neeraj  and
      Garraghan, Peter",
    editor = "Derczynski, Leon  and
      Novikova, Jekaterina  and
      Chen, Muhao",
    booktitle = "Proceedings of the The First Workshop on LLM Security (LLMSEC)",
    month = aug,
    year = "2025",
    address = "Vienna, Austria",
    publisher = "Association for Computational Linguistics",
    url = "https://aclanthology.org/2025.llmsec-1.8/",
    pages = "101--114",
    ISBN = "979-8-89176-279-4"
}

@inproceedings{markov2023holistic,
  title={A holistic approach to undesired content detection in the real world},
  author={Markov, Todor and Zhang, Chong and Agarwal, Sandhini and Nekoul, Florentine Eloundou and Lee, Theodore and Adler, Steven and Jiang, Angela and Weng, Lilian},
  booktitle={Proceedings of the AAAI conference on artificial intelligence},
  volume={37},
  number={12},
  pages={15009--15018},
  year={2023}
}

\clearpage
\newpage
\onecolumn
\raggedbottom
\appendix
\section*{Appendix}
\label{sec:appendix}
\renewcommand{\thesubsection}{A.\arabic{subsection}}

\setcounter{figure}{0}
\setcounter{table}{0}
\renewcommand{\thetable}{A.\arabic{table}}
\renewcommand{\thefigure}{A.\arabic{figure}}

\subsection{The Challenges of Single-Objective Techniques Against Guardrail Systems}

\begin{figure}[H]
    \centering
    \includegraphics[width=\linewidth]{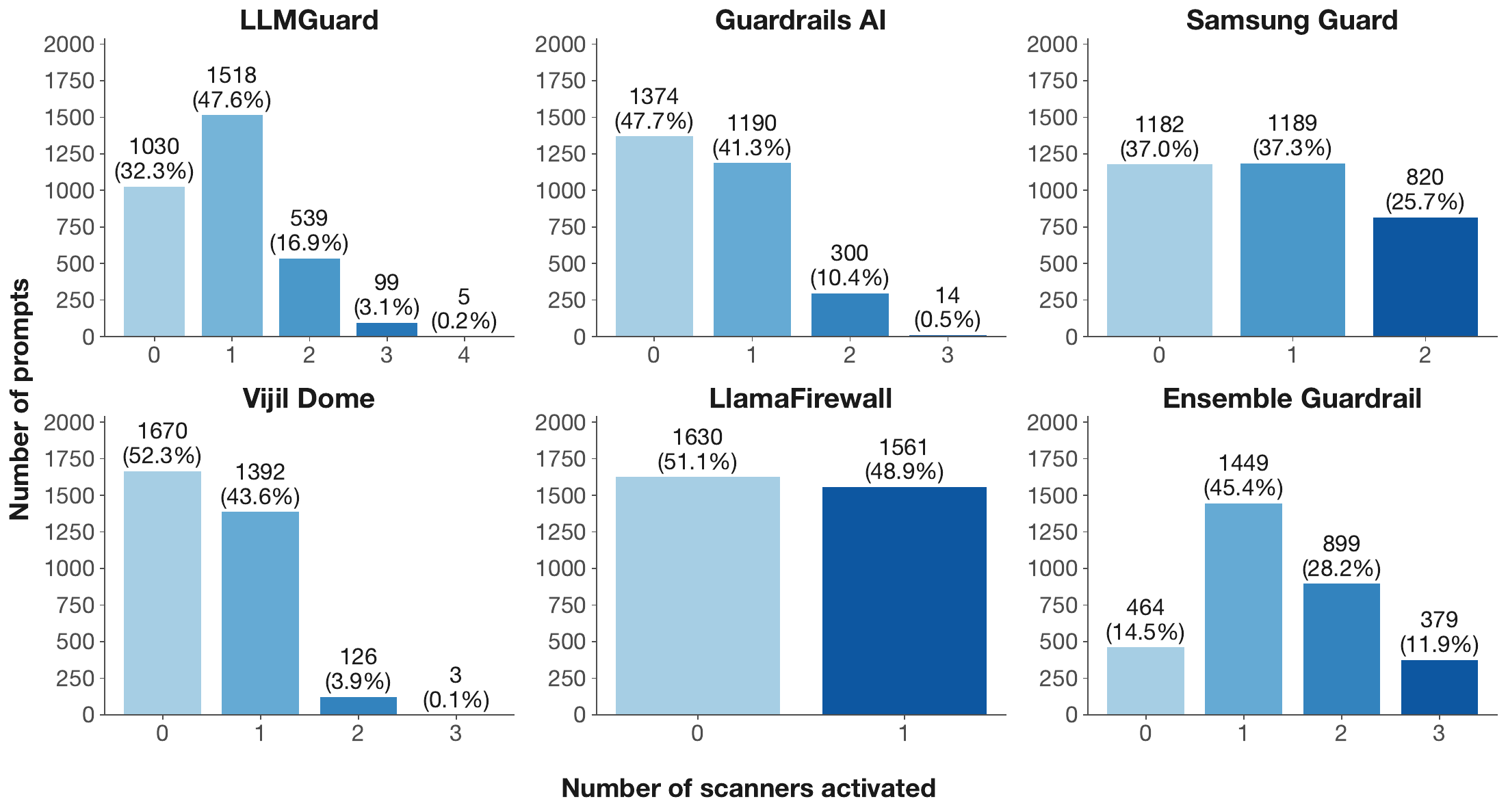}
    \caption{\textbf{Number of Scanners Activated per Prompt, Across all Targets.} Distribution of how many scanners simultaneously flag each of the 3,191 PIGuard candidate prompts, per target guardrail system.}
    \label{fig:scanners_fired_combined}
\end{figure}

\newpage
\subsection{Experimental Setup}
\begin{table}[H]
\centering
\small
\renewcommand{\arraystretch}{1.2}
\caption{\textbf{Perturbation Techniques.} All 11 perturbation techniques used in Exhaustive and BRANCH.}
\label{tab:bypass_techniques}
\begin{tabularx}{\textwidth}{l X}
\toprule
\textbf{Evasion Attack} & \textbf{Description} \\
\midrule
TextFooler \citep{textfooler} & Words with the highest importance ranking are replaced with suitable replacement words with similar semantic meaning. \\
DeepWordBug \citep{deepwordbug} & Character-level transformations are applied to the highest-ranked tokens to minimise distance of the perturbation. \\
TextBugger \citep{textbugger} & Generates utility-preserving adversarial text against black-box and white-box classification systems. \\
PWWS \citep{pwws} & Probability Weighted Word Saliency (PWWS) ranks word importance using word saliency and classification probability. \\
Pruthi \citep{pruthi} & Generates perturbations in the form of adversarial spelling mistakes via removing or swapping characters. \\
PSO \citep{pso} & Combines word substitution with particle swarm optimization (PSO) to search for adversarial candidates. \\
BAE \citep{bae} & Contextual perturbations from a BERT-MLM masked model by replacing and inserting masked tokens in the prompt. \\
CLARE \citep{clare} & Uses a pretrained masked language model to replace, insert, and merge tokens via a greedy search, constrained by sentence-embedding similarity. \\
IGA \citep{iga} & An improved genetic algorithm (GA) that swaps words with their embedding nearest-neighbours. \\
Faster Alzantot \citep{fasteralzantot} & A faster variant of Alzantot's genetic algorithm (GA) word-embedding substitution, using a smaller language model. \\
Kuleshov \citep{kuleshov} & Swaps words based upon opposite embeddings. \\
\bottomrule
\end{tabularx}
\end{table}
\newpage

\begin{table}[H]
\centering
\small
\setlength{\tabcolsep}{4pt}
\renewcommand{\arraystretch}{1.15}
\caption{\textbf{Experiment Target Detector Categories.} Highlights what each guardrail system within our experimentation is designed to detect and block. Categories vary across targets based on capability and implemented scanners.}
\label{tab:guardrail_detector_types}
\begin{tabular}{lcccccc}
\toprule
\textbf{Detector category} & \rotatebox{60}{LLM Guard} & \rotatebox{60}{Vijil Dome} & \rotatebox{60}{Guardrails AI} & \rotatebox{60}{Samsung Guard} & \rotatebox{60}{LlamaFirewall} & \rotatebox{60}{Ensemble Guardrail} \\
\midrule
Prompt Injection    & \cmark & \cmark & \xmark & \xmark & \cmark & \cmark \\
Jailbreak           & \xmark & \xmark & \cmark & \cmark & \cmark & \cmark \\
Toxicity            & \cmark & \cmark & \cmark & \xmark & \xmark & \cmark \\
Gibberish           & \cmark & \xmark & \cmark & \xmark & \xmark & \cmark \\
Harm                & \xmark & \xmark & \xmark & \cmark & \xmark & \xmark \\
Moderation          & \xmark & \xmark & \xmark & \cmark & \xmark & \xmark \\
Adversarial Suffix  & \cmark & \cmark & \xmark & \xmark & \xmark & \xmark \\
Malicious Code      & \xmark & \xmark & \xmark & \xmark & \cmark & \xmark \\
Invisible Text      & \cmark & \cmark & \xmark & \xmark & \xmark & \xmark \\
Non-English Language & \cmark & \xmark & \xmark & \xmark & \xmark & \xmark \\
Code                & \cmark & \xmark & \xmark & \xmark & \xmark & \xmark \\
\midrule
Total Categories & 7 & 4 & 3 & 3 & 3 & 4 \\
\bottomrule
\end{tabular}
\end{table}

\subsection{Query Efficiency and Technique Composition}
\label{sec:appendix_query_efficiency}

\begin{figure}[H]
    \centering
    \includegraphics[width=\linewidth]{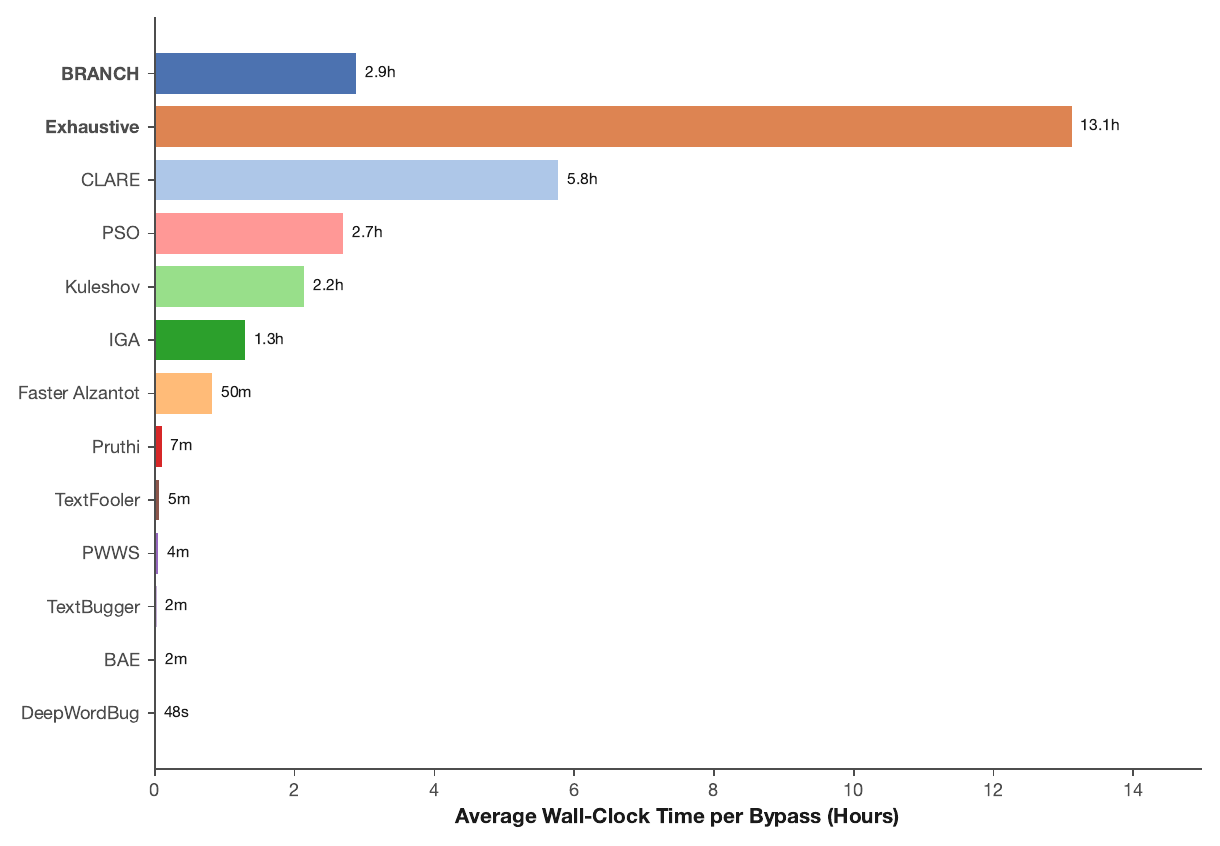}
    \caption{\textbf{Average Wall-Clock Time by Method.} Average wall-clock time per bypass job for BRANCH, Exhaustive, and the 11 individual techniques, averaged across all 6 target guardrail systems.}
    \label{fig:walltime}
\end{figure}

\begin{figure}[H]
    \centering
    \includegraphics[width=0.8\linewidth]{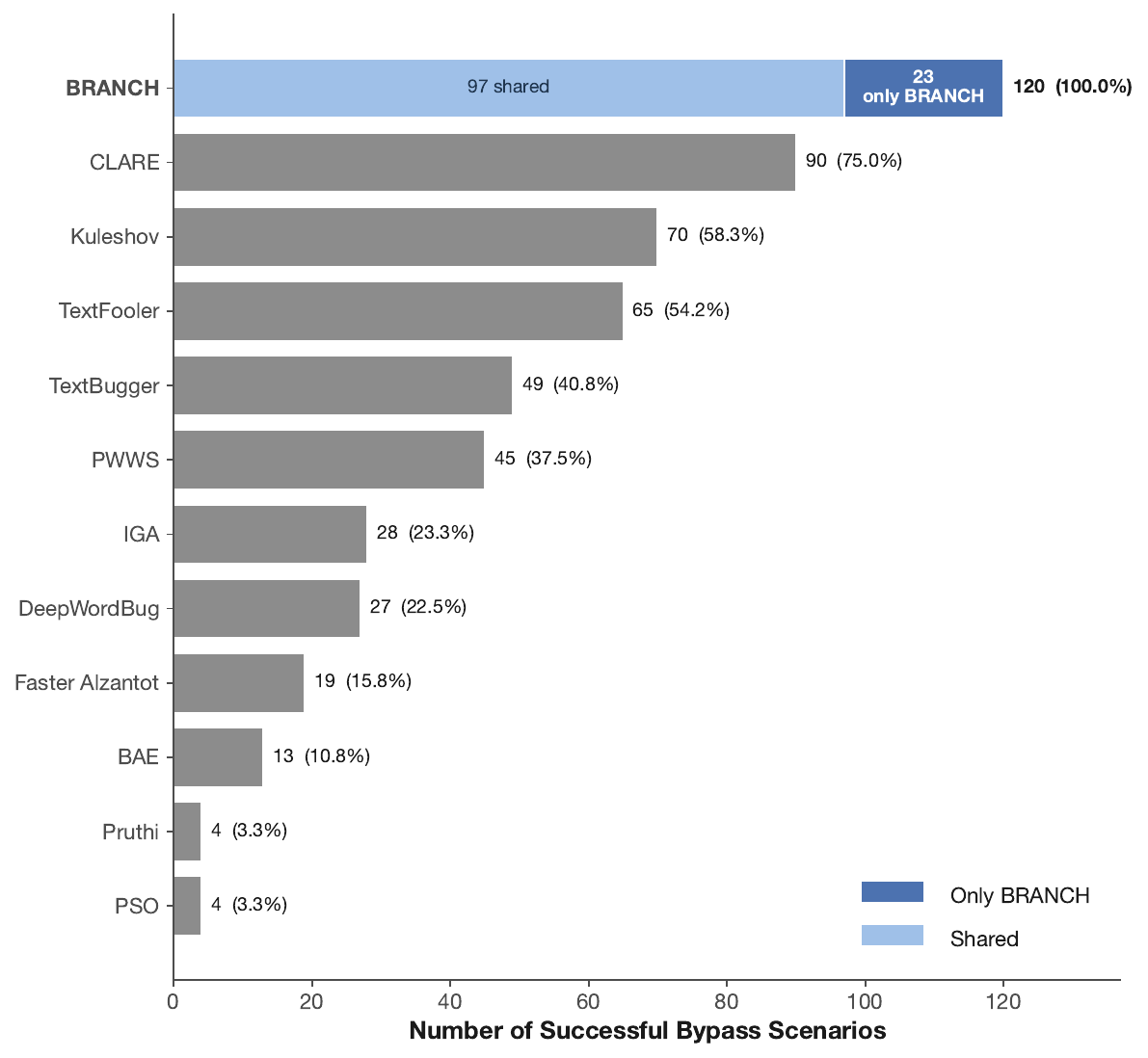}
    \caption{\textbf{BRANCH vs Individual Techniques Coverage.} Per-prompt bypass coverage, showing the effectiveness of each technique across different guardrail systems.}
    \label{fig:coverage_head_to_head}
\end{figure}

\subsection{Bypass Similarity}
\label{sec:appendix_similarity_per_target}

\begin{table}[H]
\centering
\footnotesize
\setlength{\tabcolsep}{3.5pt}
\renewcommand{\arraystretch}{1.15}
\caption{\textbf{Semantic Preservation and Attack Success Rate per Technique and Target.} $n$ is the number of successful bypasses the statistics are computed over, \textit{Pres.} the average semantic preservation $\pm$ 1 standard deviation, and the ASR over the 20 seed prompts.}
\label{tab:similarity_semantic_values}
\begin{adjustbox}{max width=\textwidth}
\begin{tabular}{l *{6}{ccc}}
\toprule
 & \multicolumn{3}{c}{\textbf{Guardrails AI}} & \multicolumn{3}{c}{\textbf{LLM Guard}} & \multicolumn{3}{c}{\textbf{LlamaFirewall}} & \multicolumn{3}{c}{\textbf{Samsung Guard}} & \multicolumn{3}{c}{\textbf{Vijil Dome}} & \multicolumn{3}{c}{\textbf{Ensemble Guardrail}} \\
\cmidrule(lr){2-4} \cmidrule(lr){5-7} \cmidrule(lr){8-10} \cmidrule(lr){11-13} \cmidrule(lr){14-16} \cmidrule(lr){17-19}
\textbf{Method} & $n$ & Pres. & ASR & $n$ & Pres. & ASR & $n$ & Pres. & ASR & $n$ & Pres. & ASR & $n$ & Pres. & ASR & $n$ & Pres. & ASR \\
\midrule
\textbf{BRANCH} & 68 & 0.65\,$\pm$\,0.17 & 100 & 75 & 0.86\,$\pm$\,0.10 & 100 & 97 & 0.81\,$\pm$\,0.12 & 100 & 56 & 0.83\,$\pm$\,0.11 & 100 & 65 & 0.81\,$\pm$\,0.11 & 100 & 53 & 0.72\,$\pm$\,0.16 & 100 \\
\textbf{Exhaustive} & 62 & 0.81\,$\pm$\,0.13 & 90 & 78 & 0.87\,$\pm$\,0.10 & 85 & 112 & 0.82\,$\pm$\,0.11 & 95 & 75 & 0.85\,$\pm$\,0.11 & 95 & 78 & 0.84\,$\pm$\,0.10 & 90 & 6 & 0.76\,$\pm$\,0.09 & 30 \\
\midrule
BAE & 3 & 0.90\,$\pm$\,0.05 & 15 & 2 & 0.95\,$\pm$\,0.02 & 10 & 3 & 0.90\,$\pm$\,0.09 & 15 & 2 & 0.97\,$\pm$\,0.03 & 10 & 3 & 0.88\,$\pm$\,0.11 & 15 & -- & -- & -- \\
CLARE & 18 & 0.73\,$\pm$\,0.15 & 90 & 17 & 0.89\,$\pm$\,0.08 & 85 & 17 & 0.86\,$\pm$\,0.06 & 85 & 19 & 0.91\,$\pm$\,0.06 & 95 & 14 & 0.89\,$\pm$\,0.09 & 70 & 5 & 0.75\,$\pm$\,0.10 & 25 \\
DeepWordBug & 5 & 0.79\,$\pm$\,0.14 & 25 & 5 & 0.83\,$\pm$\,0.14 & 25 & 9 & 0.74\,$\pm$\,0.19 & 45 & 3 & 0.65\,$\pm$\,0.19 & 15 & 5 & 0.81\,$\pm$\,0.09 & 25 & -- & -- & -- \\
Faster Alzantot & 2 & 0.94\,$\pm$\,0.02 & 10 & 3 & 0.91\,$\pm$\,0.10 & 15 & 6 & 0.79\,$\pm$\,0.08 & 30 & 5 & 0.89\,$\pm$\,0.03 & 25 & 3 & 0.84\,$\pm$\,0.09 & 15 & -- & -- & -- \\
IGA & 4 & 0.84\,$\pm$\,0.09 & 20 & 5 & 0.85\,$\pm$\,0.07 & 25 & 8 & 0.85\,$\pm$\,0.07 & 40 & 6 & 0.88\,$\pm$\,0.05 & 30 & 5 & 0.86\,$\pm$\,0.09 & 25 & -- & -- & -- \\
Kuleshov & 8 & 0.90\,$\pm$\,0.06 & 40 & 14 & 0.86\,$\pm$\,0.12 & 70 & 17 & 0.82\,$\pm$\,0.13 & 85 & 15 & 0.84\,$\pm$\,0.08 & 75 & 15 & 0.86\,$\pm$\,0.08 & 75 & 1 & 0.81\,$\pm$\,0.00 & 5 \\
Pruthi & 1 & 0.92\,$\pm$\,0.00 & 5 & 2 & 0.94\,$\pm$\,0.03 & 10 & 1 & 0.94\,$\pm$\,0.00 & 5 & -- & -- & -- & -- & -- & -- & -- & -- & -- \\
PSO & -- & -- & -- & 2 & 0.95\,$\pm$\,0.02 & 10 & -- & -- & -- & 1 & 0.83\,$\pm$\,0.00 & 5 & 1 & 0.90\,$\pm$\,0.00 & 5 & -- & -- & -- \\
PWWS & 6 & 0.82\,$\pm$\,0.07 & 30 & 7 & 0.90\,$\pm$\,0.05 & 35 & 15 & 0.80\,$\pm$\,0.11 & 75 & 8 & 0.87\,$\pm$\,0.09 & 40 & 9 & 0.81\,$\pm$\,0.06 & 45 & -- & -- & -- \\
TextBugger & 9 & 0.79\,$\pm$\,0.11 & 45 & 8 & 0.86\,$\pm$\,0.10 & 40 & 17 & 0.81\,$\pm$\,0.10 & 85 & 7 & 0.76\,$\pm$\,0.13 & 35 & 8 & 0.83\,$\pm$\,0.09 & 40 & -- & -- & -- \\
TextFooler & 7 & 0.86\,$\pm$\,0.06 & 35 & 13 & 0.83\,$\pm$\,0.10 & 65 & 19 & 0.85\,$\pm$\,0.10 & 95 & 11 & 0.84\,$\pm$\,0.13 & 55 & 15 & 0.78\,$\pm$\,0.14 & 75 & -- & -- & -- \\
\bottomrule
\end{tabular}
\end{adjustbox}
\end{table}

\begin{figure}[H]
    \centering
    \includegraphics[width=\linewidth]{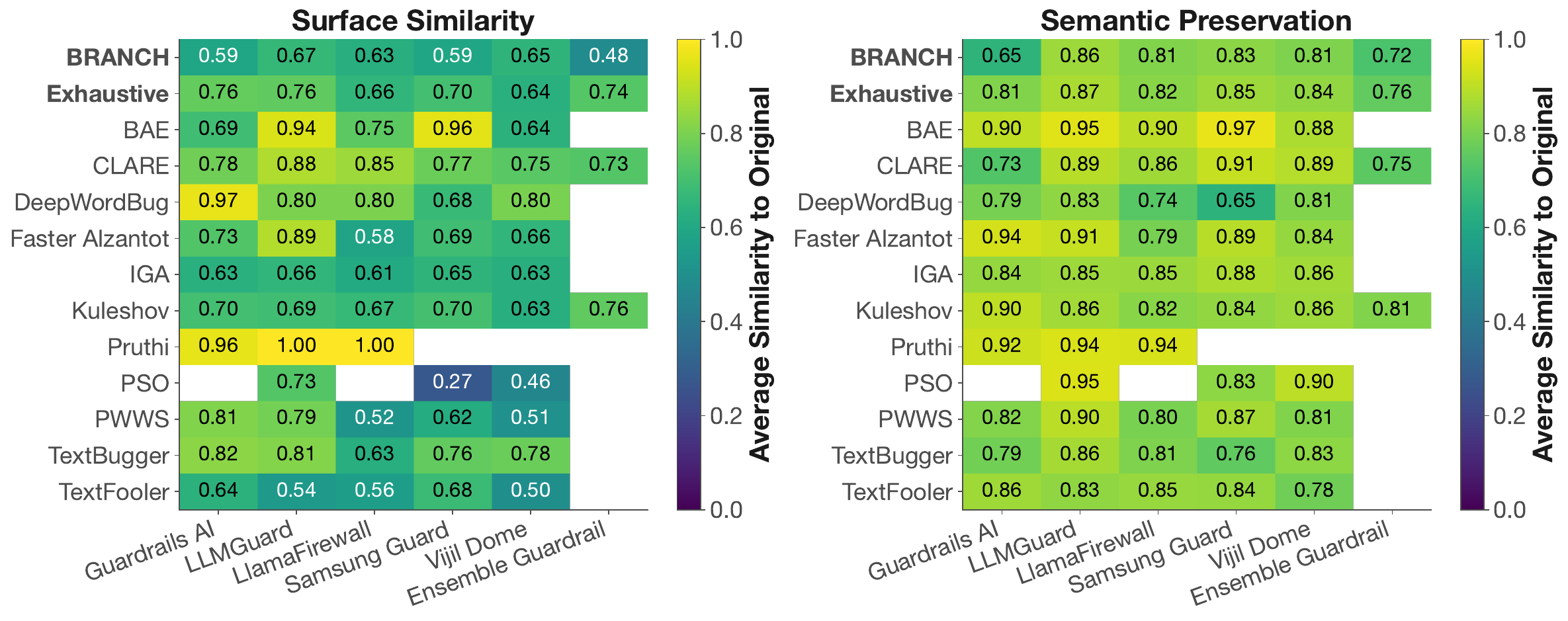}
    \caption{\textbf{Average Surface Similarity and Semantic Preservation to the Original Seed Prompt.} Exact per-target values for BRANCH, Exhaustive, and each of the 11 individual techniques.}
    \label{fig:similarity_per_target}
\end{figure}

\begin{figure}[H]
    \centering
    \includegraphics[width=\linewidth]{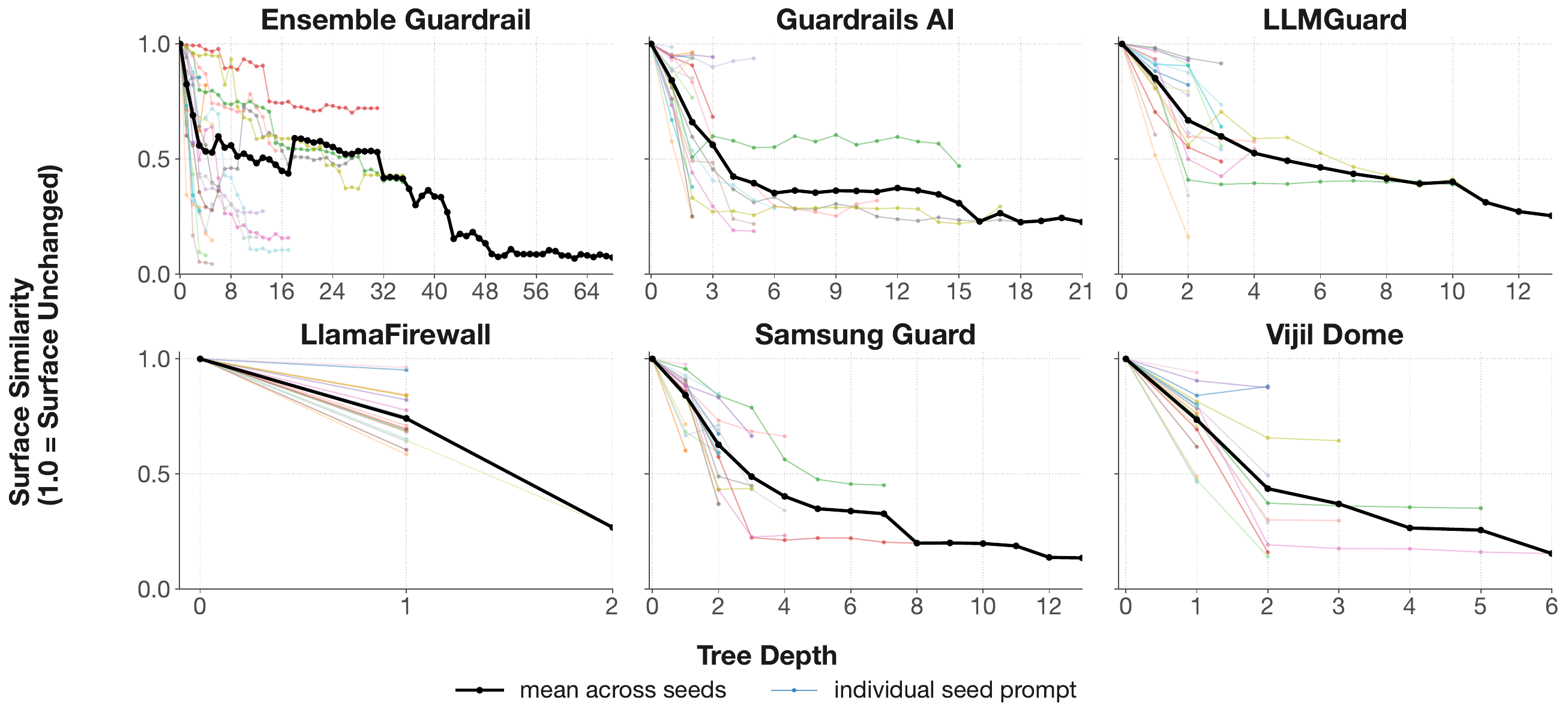}
    \caption{\textbf{BRANCH Surface Similarity.} Shows recorded surface similarity at each depth across all bypass scenarios against all targets.}
    \label{fig:depth_surface_similarity}
\end{figure}

\begin{figure}[H]
    \centering
    \includegraphics[width=\linewidth]{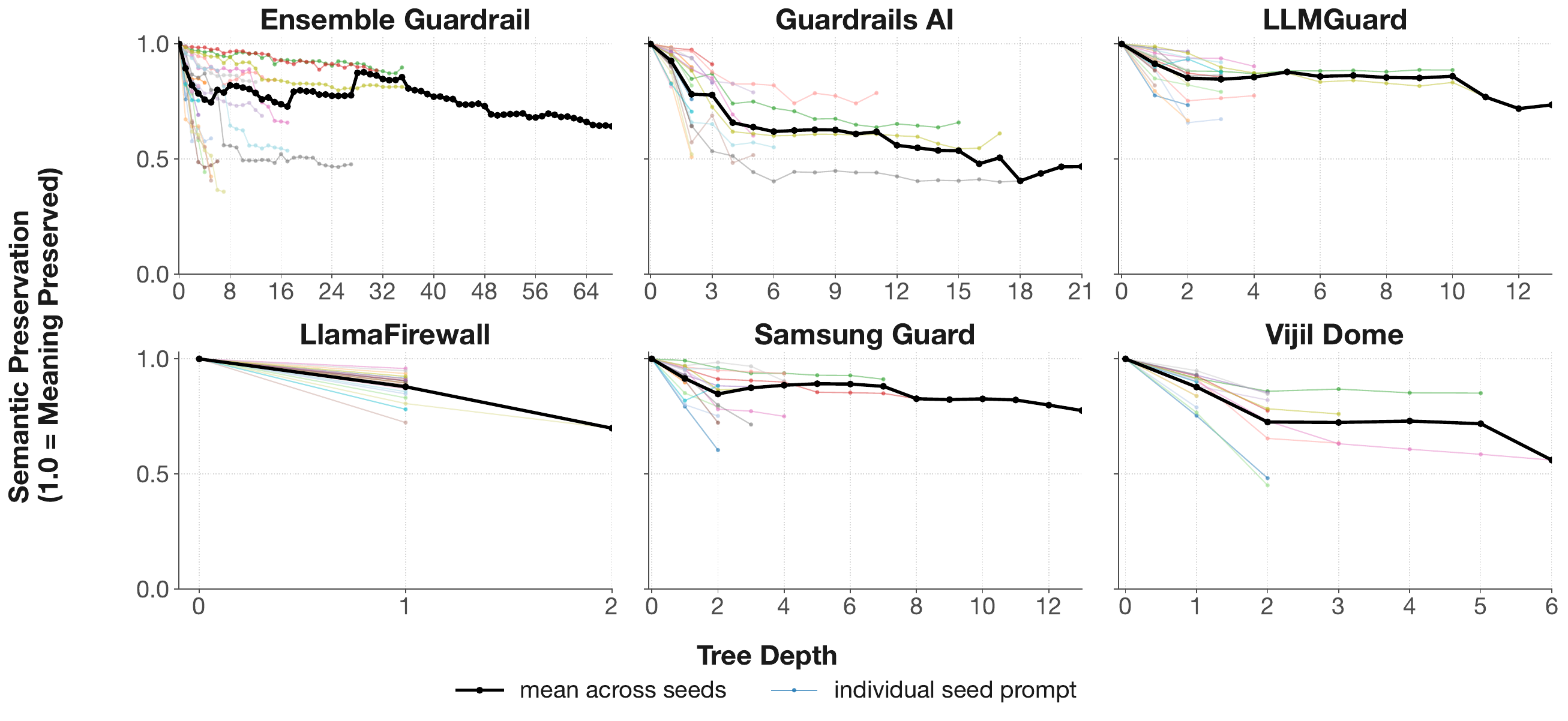}
    \caption{\textbf{BRANCH Semantic Preservation.} Shows recorded semantic preservation at each depth across all bypass scenarios against all targets.}
    \label{fig:depth_semantic_preservation}
\end{figure}

\subsection{Transferability}
\label{sec:appendix_transferability}

\begin{figure}[H]
    \centering
    \includegraphics[width=\linewidth]{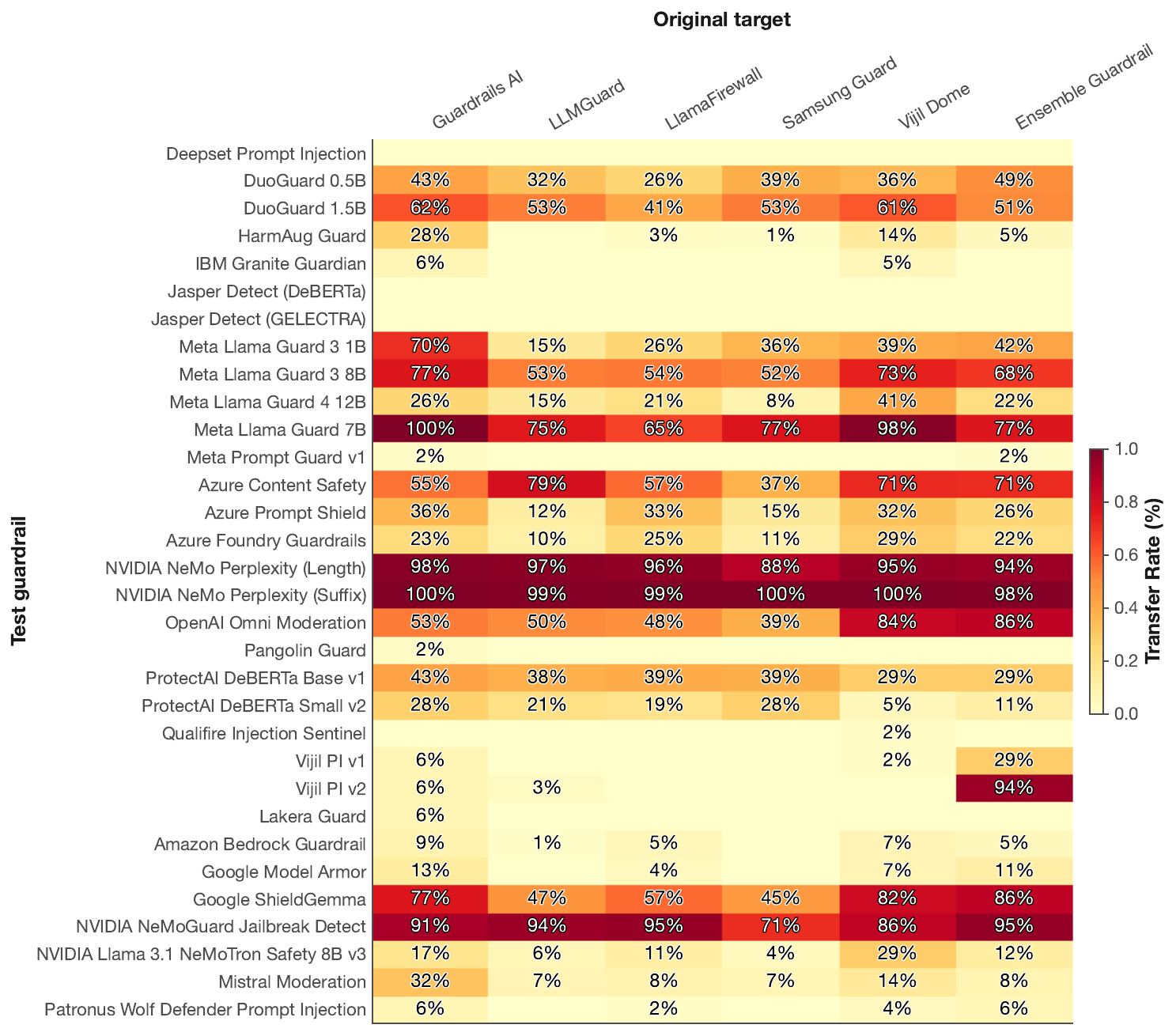}
    \caption{\textbf{Transferability Heatmap for BRANCH.} Shows percentage of generated bypasses from the original target (x) transferring to an unseen guardrail (y).}
    \label{fig:transfer_heatmap_branch}
\end{figure}

\end{document}